\documentclass[submission,copyright,creativecommons]{eptcs}
\providecommand{\event}{SOS 2007} % Name of the event you are submitting to
\usepackage{underscore}           % Only needed if you use pdflatex.
\usepackage{soul}
\usepackage{url}
\usepackage[utf8]{inputenc}
\usepackage[small]{caption}
\usepackage{graphicx}
\usepackage{amsmath}
\usepackage{booktabs}
\usepackage{paralist}
\usepackage{amsthm}
\usepackage{amssymb}
\usepackage{subcaption}
\usepackage{xcolor}
\newcommand{\LTLf}{{\sc ltl}$_f$}
\newcommand{\ltlf}{{\sc ltl}$_f$}
\newcommand{\ltl}{{\sc ltl}}
\newcommand{\Syft}{\textsc{Syft}}
\newcommand{\Mona}{\textsc{Mona}}
\newcommand{\Spot}{\textsc{Spot}}
\DeclareMathAlphabet{\mathcal}{OMS}{cmsy}{m}{n}
\newcommand{\X}{\mathcal{X}}
\newcommand{\Y}{\mathcal{Y}}
\newcommand{\Z}{\mathcal{Z}}
\newcommand{\Prop}{\mathcal{P}}

\newtheorem{definition}{Definition}
\newtheorem{theorem}{Theorem}
\allowdisplaybreaks

\title{LTLf Synthesis under Partial Observability: \\ From Theory to Practice}
\author{Lucas M. Tabajara
\institute{Rice University\\ Houston, USA}
\email{lucasmt@rice.edu}
\and
Moshe Y. Vardi
\institute{Rice University\\ Houston, USA}
\email{vardi@cs.rice.edu}
}
\def\titlerunning{LTLf Synthesis under Partial Observability: From Theory to Practice}
\def\authorrunning{L.M. Tabajara \& M.Y. Vardi}
\begin{document}
\maketitle

\begin{abstract}
{\ltl} synthesis is the problem of synthesizing a reactive system from a formal specification in Linear Temporal Logic. The extension of allowing for partial observability, where the system does not have direct access to all relevant information about the environment, allows generalizing this problem to a wider set of real-world applications, but the difficulty of implementing such an extension in practice means that it has remained in the realm of theory. Recently, it has been demonstrated that restricting {\ltl} synthesis to systems with finite executions by using {\ltl} with finite-horizon semantics ({\ltlf}) allows for significantly simpler implementations in practice. With the conceptual simplicity of {\ltlf}, it becomes possible to explore extensions such as partial observability in practice for the first time. Previous work has analyzed the problem of {\ltlf} synthesis under partial observability theoretically and suggested two possible algorithms, one with 3EXPTIME and another with 2EXPTIME complexity. In this work, we first prove a complexity lower bound conjectured in earlier work. Then, we complement the theoretical analysis by showing how the two algorithms can be integrated in practice into an established framework for {\ltlf} synthesis. We furthermore identify a third, MSO-based, approach enabled by this framework. Our experimental evaluation reveals very different results from what the theory seems to suggest, with the 3EXPTIME algorithm often outperforming the 2EXPTIME approach. Furthermore, as long as it is able to overcome an initial memory bottleneck, the MSO-based approach can often outperforms the others.
\end{abstract}

\section{Introduction}

{\ltl} synthesis~\cite{PnueliR89} is the problem of automatically generating a reactive system from a high-level specification of its behavior described in Linear Temporal Logic ({\ltl})~\cite{Pnueli77}. Since its introduction~\cite{PnueliR89}, this problem has become a prominent area of research in formal methods, with a number of {\ltl}-synthesis tools~\cite{Ehlers11,BohyBFJR12,FaymonvilleFT17,MeyerSL18} being developed over the years despite the problem's 2EXPTIME-completeness and the fact that algorithms often rely on complex operations such as determinization of $\omega$-automata~\cite{FogartyKVW13} and parity-game solving~\cite{Zielonka98}. A line of follow-up work has focused on extending and generalizing the problem, 
such as allowing for partial observability (incomplete information)~\cite{KupfermanV97}. Because this adds an additional layer of complexity over the already-complex {\ltl}-synthesis algorithms, however, this extension has resisted practical implementation and therefore have mostly remained in the theoretical realm.

Recent times have seen interest in a finite-horizon variant of {\ltl} called Linear Temporal Logic over finite traces ({\ltlf})~\cite{GiacomoV13}. {\ltlf} can be used to describe reactive systems with finite executions, making it relevant for example in the area of robotics~\cite{HeWKV19}, and {\ltlf} synthesis is closely connected to planning in AI~\cite{CamachoBM19}. Despite having the same doubly-exponential complexity as the synthesis problem for {\ltl} over infinite traces~\cite{GiacomoV15}, the appeal of {\ltlf} synthesis is that it requires much simpler algorithms, which translates into better practical performance. {\ltlf} specifications can be translated into finite automata, which are much easier to determinize and minimize than $\omega$-automata, and can then be synthesized by playing a reachability game over the state space of the automaton~\cite{GiacomoV15}, rather than a more complex type of game such as a parity game. Thus, practical tools for {\ltlf} synthesis have already started being developed~\cite{ZhuTLPV17,CamachoBMM18} which compare very favorably with existing {\ltl}-synthesis tools. These successes suggest that extensions of the synthesis problem that have resisted implementation in the infinite setting might also now have the potential to be realized in the finite-horizon case.

The extension that this paper focuses on is \emph{synthesis under partial observability}, also called \emph{synthesis with incomplete information}~\cite{KupfermanV97}. This generalization introduces \emph{unobservable inputs}, which are propositions on which the specification depends but whose value is unknown to the system. This variant of the problem can thus model scenarios in which the system does not have access to all relevant information about the environment at all times; for example, a robot that is only able to sense its local vicinity. The extension of {\ltlf} synthesis from full to partial observability was first investigated in~\cite{GiacomoV16}, which presents two approaches for this problem: one based on the construction of a belief-states space and a projection-based approach. Although the belief-states approach leads to a 3EXPTIME complexity, the projection-based approach is 2EXPTIME, matching the complexity of both synthesis under full observability and planning under partial observability~\cite{Rintanen04a}, both of which this problem generalizes. 

The analysis in~\cite{GiacomoV16} was theoretical, but with the subsequent development of practical tools for {\ltlf} synthesis we can now investigate how practical concerns may affect these results. Although the theoretical analysis suggests a clear advantage to the projection-based approach, there have been examples, when dealing with automata, of worst-case exponential gaps not manifesting in practice. For example, NFA often become smaller when determinized and minimized, even though in the worst case the minimal DFA may be exponentially larger~\cite{TabakovRV12}.
Thus, to understand how to best solve the problem of synthesis under partial observability it is necessary to observe the performance of the algorithms in practice.

This works offers the following three contributions, which complement the results of~\cite{GiacomoV16}:
First, we prove the conjecture from~\cite{GiacomoV16} that synthesis under partial observability for NFA specifications is 2EXPTIME-hard (and therefore 2EXPTIME-complete). This result completes the landscape of theoretical complexity presented in that work, which had proved tight bounds for DFA, AFW, and logical specifications. Second, we investigate how the two approaches for {\ltlf} synthesis under partial observability discussed in~\cite{GiacomoV16} can be implemented in practice within the {\Syft} framework~\cite{ZhuTLPV17}, which currently represents the state of the art in {\ltlf} synthesis. Benefiting from {\Syft}'s use of a symbolic synthesis algorithm, we implement the two approaches symbolically, potentially avoiding an exponential memory blowup. We additionally propose a third, alternative approach for the problem that naturally emerges from {\Syft}'s use of the tool {\Mona}~\cite{HenriksenJJKPRS95} to convert from monadic second-order logic (MSO)~\cite{Buchi1960} to DFA.

Finally, we implement the three approaches within {\Syft} and evaluate their performance, thus complementing the theoretical analysis from~\cite{GiacomoV16} with an empirical evaluation. To the best of our knowledge, this is the first instance of algorithms for temporal synthesis under partial observability being implemented in practice and evaluated empirically. Our evaluation reveals that the story is more nuanced than the theoretical analysis would lead us to believe. While in terms of worst-case complexity there is an exponential gap between the belief-states and projection-based approaches, this gap does not necessarily appear in practice, and in fact the projection-based approach turns out to be in many cases outperformed by the belief-states approach due to the latter producing a more efficient symbolic representation. We also find that while the MSO-based approach leads to significantly larger automata initially and is more likely to run out of memory during automaton construction, if this hurdle is overcome, then synthesis tends to be more efficient than the other approaches. This suggests that the MSO approach may be a promising option for {\ltlf} synthesis under partial observability, and furthermore motivates improving automata-construction algorithms from MSO formulas.

\section{Preliminaries}

\paragraph{Linear Temporal Logic over Finite Traces}

Linear Temporal Logic over finite traces, i.e. {\ltlf}~\cite{GiacomoV13} extends propositional logic with finite-horizon temporal operators.
{\ltlf} is a variant of Linear Temporal Logic, or {\ltl}~\cite{Pnueli77}, with the difference that {\ltlf} is interpreted over finite traces, rather than infinite traces as in {\ltl}. Given a set of propositions $\Prop$, the syntax of {\ltlf} is identical to {\ltl}, and defined as:
$$\varphi ::= \top\ |\ \bot\ |\ p \in \Prop\ |\ (\neg \varphi)\ |\  (\varphi_1\wedge\varphi_2)\ |\ (X\varphi)\ |\ (\varphi_1 U \varphi_2)$$

$\top$ and $\bot$ represent \textit{true} and \textit{false} respectively. 
$X$ for ``Next" and $U$ for ``Until" are temporal operators. Other operators can be written in terms of those.
A \textit{trace} $\rho = \rho[0],\rho[1],\ldots$ is a sequence of propositional interpretations (sets) $\rho[i]\in 2^\Prop$. Intuitively, $\rho[i]$ is interpreted as the set of propositions which are $true$ at instant $i$.
Trace $\rho$ is an \textit{infinite} trace if $|\rho| = \infty$, denoted as $\rho\in (2^\Prop)^{\omega}$; 
otherwise $\rho$ is a \textit{finite} trace, denoted as $\rho\in (2^\Prop)^{*}$. We assume standard semantics from~\cite{GiacomoV13}.

An {\ltlf} formula can be represented by an automaton over finite words that accepts a trace if and only if that trace satisfies the formula.
A \emph{nondeterministic finite automaton} (NFA) is a tuple $A = (\Sigma, S, s_0, \delta, F)$, where $\Sigma$ is the alphabet, $S$ is the state space, $s_0 \in S$ is the initial state, $\delta : S \times \Sigma \rightarrow 2^S$ is the (nondeterministic) transition function and $F \subseteq S$ is the set of accepting states. If the transition function $\delta$ is such that $|\delta(s, \sigma)| = 1$ for all $s \in S$ and $\sigma \in \Sigma$, then we say that $A$ is a \emph{deterministic finite automaton} (DFA) and we simplify the signature of $\delta$ to $\delta : S \times \Sigma \rightarrow S$.
In the case of finite automata obtained from an {\ltlf} formula, the alphabet is comprised of interpretations to the propositions of the formula, i.e. $\Sigma = 2^\Prop$. In this case, it is often useful to represent the transition function symbolically using Binary Decision Diagrams (BDDs)~\cite{Bryant86} and similar data structures.

\paragraph{{\ltlf} Synthesis}

The \emph{full-observability} version of the problem of \emph{{\ltlf} synthesis}~\cite{GiacomoV15} is defined as follows:

\begin{definition}[{\LTLf} Synthesis]
	Let $\varphi$ be an {\LTLf} formula over $\Prop$ and $\X$, $\Y$ be two disjoint sets of propositions such that $\X \cup \Y = \Prop$. $\X$ is the set of \emph{input variables} and $\Y$ is the set of \emph{output variables}. $\varphi$ is \emph{realizable} with respect to $\langle\X, \Y\rangle$ if there exists a strategy $f: (2^{\X})^*\rightarrow 2^{\Y}$, such that for an arbitrary infinite sequence $\pi = X_0,X_1,\ldots\in (2^{\X})^{\omega}$ of propositional interpretations over $\X$, we can find $k\geq 0$ such that the finite trace $\rho = (X_0\cup f(\epsilon)), (X_1\cup f(X_0)),\ldots , (X_k\cup f(X_0,X_1,\ldots,X_{k-1}))$ satisfies $\varphi$.
\end{definition} 

Intuitively, {\LTLf} synthesis can be thought of as a game between two players: the \emph{environment}, who controls the input variables, and the \emph{system}, who controls the output variables. Solving the synthesis problem means synthesizing a strategy $f$ for the system such that no matter how the environment behaves, the combined behavior trace of both players satisfy the logical specification $\varphi$~\cite{GiacomoV15}.

In~\cite{GiacomoV15} the authors introduce an algorithm for {\ltlf} synthesis based on a reduction to a DFA game. The current state of the art for solving this problem is based on a symbolic version of this algorithm, proposed in~\cite{ZhuTLPV17}. Refer to those papers for details. In~\cite{GiacomoV16} the authors extend the problem of {\ltlf} synthesis to the setting of \emph{partial observability}, where the system does not have complete information about what happens in the environment. This situation is modeled by partitioning the set of input propositions $\X$ into $Obs$, the set of \emph{observable} propositions, and $Unobs$, the set of \emph{unobservable} propositions. When deciding on an action, the system can only base its decision on the observable inputs. Therefore, this variant of the problem asks for a strategy $f : (2^{Obs})^* \rightarrow 2^\Y$ such that for all infinite sequences $X_0, X_1, \ldots \in (2^\X)^\omega$, a finite trace $(X_0 \cup f(\epsilon)), (X_1 \cup f(X_0|_{Obs}), \ldots, (X_k \cup f(X_0|_{Obs}, X_1|_{Obs}, \ldots, X_{k-1}|_{Obs}))$ satisfies the specification, for some $k \geq 0$. Both the full- and partial-observability versions are 2EXPTIME-complete~\cite{GiacomoV15,GiacomoV16}.
\section{Partial Observability for NFA Specifications}

In addition to proving 2EXPTIME-completeness of {\ltlf} synthesis under partial observability, \cite{GiacomoV16} also analyzed the complexity of the problem when starting already from automaton specifications. The problem was proved to be EXPTIME-complete from a DFA specification, and 2EXPTIME-complete from a specification given as a alternating finite-word automaton (AFW). For NFA specifications, the problem was shown to be in 2EXPTIME, but no lower bound was proved, although the authors conjectured that it was 2EXPTIME-complete.
In this section we present a sketch of a proof that this conjecture is correct, and synthesis under partial observability from NFA specifications is indeed 2EXPTIME-complete. We prove the lower bound by simulating an alternating Turing machine that uses at most exponential space. As it is known that $AEXPSPACE = 2EXPTIME$~\cite{ChandraKS81}, this proves that the problem is 2EXPTIME-complete.
The reduction uses a technique of modeling configurations of the Turing machine using the alphabet of the automaton (see~\cite{VardiS85,SistlaVW85,Rosner91}). For ease of exposition, we first describe a reduction to an NFA with polynomial number of states but with an exponential-size alphabet. Later we explain how to modify the reduction to use a polynomial alphabet.

Let $M = (Q, \Gamma, \delta, q_0, g)$ be an alternating Turing machine (ATM)~\cite{ChandraKS81} that requires space at most $2^{cn}$, where $n$ is the size of the input and $c$ is a constant. $Q$ is the set of states, $\Gamma$ the tape alphabet, $\delta : Q \times \Gamma \rightarrow \mathcal{P}(Q \times \Gamma \times \{L, R\})$ is the transition function, $q_0 \in Q$ is the initial state and $g : Q \rightarrow \{\forall, \exists, accept, reject\}$ indicates whether a state is universal, existential, accepting or rejecting. Transitions $(q', \gamma', d) \in \delta(q, \gamma)$ indicate the next state $q'$ of the machine, the symbol $\gamma'$ to write on the tape, and the direction $d$ to move the head. Computations of an ATM can be seen as a game between a \emph{universal} and an \emph{existential} player. Which transition in $\delta(q, \gamma)$ is taken is chosen by the universal player if $g(q) = \forall$ and the existential player if $g(q) = \exists$. The machine accepts if the existential player has a strategy to reach an accepting state.

For simplicity, assume $\Gamma = \{0, 1, \#\}$, where $\#$ is the blank symbol. Let $x = x_1 \ldots x_n \in \Gamma^*$ be an input string, which starts out on the tape. We construct an instance of the problem of NFA synthesis under partial observability that is realizable if and only if $M$ accepts $x$. This instance is given by an NFA $N = (\Sigma, S, \Delta, s_0, F)$, with alphabet $\Sigma = Obs \times Unobs \times Out$, where $Obs$ is the set of observable inputs, $Unobs$ is the set of unobservable inputs and $Out$ is the set of outputs. Note that in this case $Obs$, $Unobs$ and $Out$ are sets of symbols rather than of propositions, but if desired each can be encoded using a logarithmic number of propositions.

\vspace{-2.5mm}
\paragraph{Simulating ATM Computations}

In the reduction, the environment and the system take the roles of universal and existential players, respectively. We define $Obs = \{1, \ldots, m_\forall\}$, where $m_\forall$ is the highest branching factor of a universal state in $M$ (i.e. $m_\forall = \max \{ |\delta(q, \gamma)| \mid g(q) = \forall \}$). If the current state is universal, the environment player uses the observable inputs to choose a transition from $\Delta(q, \gamma)$. Likewise, $Out = \{1, \ldots, m_\exists\} \times \{1, \ldots, 2^{cn}\} \times (\Gamma \cup (Q \times \Gamma))$, where $m_\exists$ is similarly the highest branching factor of an existential state in $M$ (i.e. $m_\exists = \max \{ |\delta(q, \gamma)| \mid g(q) = \exists \}$). The first component of $Out$ is similarly used by the system player to choose a transition from an existential state. The other two components are used to encode a cell $(k, u)$, where $k$ is a counter indicating which position of the tape the cell occupies and $u$ is the contents of the cell, which are either a symbol $\gamma$ or a tuple $(q, \gamma)$ if the head of the machine is on that cell and on state $q$.

Once we have taken care of universal and existential branching, the idea of the reduction is that a trace of the NFA represents a sequence of configurations of the ATM. A configuration is given by a sequence of cells $(k, u)$ of the form $(1, u_1), (2, u_2), \ldots, (2^{cn}, u_{2^{cn}})$. After $k = 2^{cn}$, in the next step it should reset back to $1$, indicating the start of a new configuration that should follow from the previous one according to the transition function $\delta$. The NFA accepts if the trace reaches an accepting configuration.

The challenge of the reduction is to enforce that the configurations produced by the system player are consistent: that the counter $k$ increases by $1$ each time and resets after $2^{cn}$, and that one configuration follows from the previous one, and the first configuration has $x$ on the tape. If this is the case, then the trace of the NFA corresponds to an accepting computation of $M$. We cannot enforce this consistency just by storing information in the state, because this would require an exponential number of states. Instead, we use the unobservable inputs to constrain the actions of the system player. 

\vspace{-2.5mm}
\paragraph{Using Partial Observability and Nondeterminism}

We define $Unobs = \{1, \ldots, cn\} \times \{0, 1\}$. The first component $i \in \{1, \ldots, cn\}$ is only used at the first step of the trace, and represents the choice of a bit $k_i$ of the counter $k \in \{1, \ldots, 2^{cn}\}$ for the environment to monitor. At each step of the computation, the automaton will determine from the current value of $k$ what the value of $k_i$ should be at the next step, and store that in the state. If at any point the value of $k_i$ differs from the expected, the NFA rejects. Note that, since the system player does not know which bit the environment has chosen to monitor, the only way to guarantee a win is to always keep the entire counter consistent.

The second component $p \in \{0, 1\}$ is a flag that should be raised exactly twice during the computation, on two adjacent configurations (if the environment breaks this assumption, the NFA accepts). If on the two times that $p = 1$ the counter has the same value (i.e., $p$ points to the same cell both times), then the contents of the cell on the second configuration must follow from the first configuration according to the transition relation (e.g., if the head was in that cell, it must have written the correct symbol and moved away, etc.). To check for that, $N$ makes a nondeterministic guess the first time $p = 1$. If $N$ guesses that $p$ will point to a different cell, it guesses also which bit will be different between the two counters and stores that in the state. The second time $p = 1$, $N$ checks that the bits are indeed different. If it guesses that $p$ will point to the same cell, it stores in the state what the value of the cell should be in the next configuration, and checks that it is correct once $p = 1$ again. Similarly to the counter, since the system player does not know when $p$ is raised, the only way to guarantee a win is to ensure that adjacent configurations follow from one another.

\vspace{-2.5mm}
\paragraph{Polynomial State Space.}
Note that the states of $N$ must keep track of the following information:
which state $q \in Q$ the machine is in; what was the bit $i \in \{1, \ldots, cn\}$ chosen by the environment in the first step of the trace; how many times $p$ has been raised (0, 1, 2 or more); if $p$ has been raised once, how long ago that was (this configuration, last configuration, earlier than that); if $N$ has guessed that $p$ will point to different cells, what is the index $i$ and value $k_i$ of the bit that will be different; if $N$ has guessed that $p$ will point to the same cell, what is the expected content of that cell in the next configuration; the contents of the previous cell on the tape, in case $p$ is raised (the contents of a cell can be affected only by its adjacent cells).
Since each component of the state is polynomial on $M$ and $x$, the NFA has a polynomial number of states. Accepting states are those where $g(q) = accept$. For lack of space, we omit the details of the transition function.

\vspace{-2.5mm}
\paragraph{Polynomial Alphabet.}
Note that the alphabet $\Sigma$ of $N$ is polynomial except for the counter $k$ that forms the second component of $Out$. We can reduce the alphabet to polynomial size by encoding each cell $(k, u)$ over multiple time steps as a sequence $k_1, \ldots, k_{cn}, u$, where $k_i \in \{0, 1\}$ is the $i$-th bit of $k$. This requires splitting each state of the automaton into $cn + 1$ states, and also keeping track of additional information in the state (necessary, for example, to compute the next value of the bit $k_i$ being monitored by the environment). Yet, none of these changes make the state space larger than polynomial.

Therefore, the reduction from acceptance of an ATM to synthesis under partial observability from an NFA specification is polynomial.

\begin{theorem}
Synthesis under partial observability from an NFA specification is 2EXPTIME-complete.
\end{theorem}
\section{Partial-Observability Synthesis in Practice} \label{sec:approaches}

Two algorithms for {\ltlf} synthesis under partial observability were proposed in~\cite{GiacomoV16}:
a belief-states construction with worst-case 3EXPTIME complexity and a projection-based construction that achieves an optimal 2EXPTIME complexity.
In this section we show how algorithms for synthesis under partial observability can be practically implemented within the context of existing tools for {\ltlf} synthesis. We first review the {\Syft} framework~\cite{ZhuTLPV17}, which represents the state-of-the-art for {\ltlf} synthesis under full observability, combining an explicit automaton construction with symbolic BDD-based techniques for synthesizing the strategy efficiently.
Then, we introduce novel
versions of the two algorithms for partial observability that perform part of the automaton construction symbolically. This serves two purposes. First, it allows them to be easily integrated into {\Syft}'s framework, as the symbolic automata can be passed directly to the symbolic strategy computation. Second, it avoids an explicit  exponential blow-up in the automaton-construction step of the algorithms, as it avoids ever constructing the final automaton explicitly and instead directly constructs a symbolic representation. This representation tends to be much more compact and sometimes exponentially smaller. Finally, we describe a third, novel MSO-based approach that is made possible specifically by the DFA-construction approach employed by {\Syft}.

\paragraph{The {\Syft} Framework.}
{\Syft}'s synthesis approach can be summarized as follows.
    First, translate the {\ltlf} formula $\varphi$ into a formula in first-order logic $fol(\varphi)$, using the procedure described in~\cite{GiacomoV13}.
    Then, use the tool {\Mona}~\cite{HenriksenJJKPRS95} to convert $fol(\varphi)$ into a minimal DFA $A$.
    Next, convert $A$ into a symbolic-state representation over a set of state variables $\Z$, logarithmic in the number of states. Each state is implicitly encoded as an interpretation of the variables in $\Z$, and the transition relation and set of accepting states are then represented by BDDs.
    Finally, use a symbolic fixpoint algorithm to compute a winning strategy in the DFA game given by $A$.
Details of each step can be found in~\cite{ZhuTLPV17}.

\subsection{Projection-Based Construction} \label{sec:double-negation}

We start by describing the second approach from~\cite{GiacomoV16}, as the first approach can be seen as a special case of it. We can summarize this approach as follows:
\begin{inparaenum} \item construct an NFA $\bar{N}$ for $\neg \varphi$; \item project unobservable inputs from $\bar{N}$'s transition function; \item determinize $\bar{N}$ into a DFA $\bar{A}$; \item complement $\bar{A}$ into $A$. \end{inparaenum}
After the second step, $\bar{N}$ accepts those traces that can be extended by a trace of unobservable inputs such that the result violates $\varphi$. By complementing the automaton we obtain a DFA game that can be won by the system iff $\varphi$ can be realized under partial observability. This construction takes advantage of the fact that
{\ltlf} formulas are closed under negation, NFAs are closed under projection and DFAs are closed under complementation, and each of these operations can be performed in polynomial time.
Therefore, the only exponential steps are the conversions from {\ltlf} to NFA and NFA to DFA, making the entire construction doubly exponential.

The challenge in implementing this construction in the {\Syft} framework is that {\Syft} is based on {\Mona}, which translates logical formulas to DFAs, while
we need to first construct an NFA $\bar{N}$ for $\neg \varphi$. We do this in two steps.
First, we construct a minimal DFA for the \emph{reverse} of the language of $\neg \varphi$ (this DFA is guaranteed to be at most exponential in the size of the formula~\cite{ChandraKS81}). Then, we reverse this DFA by switching the initial and final states and reversing all transitions. The result is an NFA for the language of $\neg \varphi$, and this NFA is at most exponential. To construct the DFA for the reverse language, we employ a technique introduced in~\cite{ZhuPV19}, which converts an {\ltlf} formula into a Past {\ltlf} formula for its reverse language, then converts this Past {\ltlf} formula into first-order logic to give as input to {\Mona}. Besides providing theoretical guarantees that the NFA constructed is exponential at most, this approach has also performed well in our preliminary experiments against alternative approaches for NFA construction, such as using the automaton package {\Spot}~\cite{spot}.

The next three steps, particularly the determinization step, may lead to an exponential blow-up in the automaton. To mitigate this problem, we describe how to perform these steps symbolically, so that we construct a symbolic representation of the DFA directly from the explicit representation of $\bar{N}$, without ever building the state space of the DFA explicitly. This can be done because the standard subset-construction approach for determinization lends itself naturally to being performed symbolically.
Because the symbolic representation can be exponentially more compact, this construction might avoid an explicit exponential blowup. We now describe the symbolic construction.

The NFA $\bar{N} = (2^\Prop, S, \delta, s_0, F)$ is generated with transitions represented symbolically by a BDD $T_{i, j}$ for every pair of states $s_i$ and $s_j$, such that $T_{i, j}$ evaluates to $1$ under an interpretation $\sigma \in 2^\Prop$ iff $s_j \in \delta(s_i, \sigma)$. We project the unobservable propositions by simply applying a standard BDD operation of existential quantification to every $T_{i, j}$. To perform determinization symbolically, we create a state variable $z_i$ for each state $s_i$ of $\bar{N}$. Then, an interpretation $Z$ to the state variables $\Z$ represents the subset that contains exactly those states for which the corresponding variable is $true$.
The transition function is then represented by BDDs $\Delta_1, \ldots, \Delta_{|\Z|}$, where $\Delta_j = \bigvee_{z_i \in \Z} (z_i \land \exists u_1, \ldots, u_n . T_{i, j})$ for $Unobs = \{u_1, \ldots, u_n\}$. Note that $\Delta_j$ evaluates to $1$ iff $z_j$ is in the successor subset according to subset construction. The accepting states (after complementation) are also represented by a BDD $\Phi = \neg \bigvee_{s_i \in F} z_i$, which evaluates to $1$ for an interpretation $Z$ if $Z$ represents an accepting subset.
Note that the existential quantification in $\Delta_j$ and the negation in $\Phi$ come respectively from the projection and complementation steps. This final symbolic DFA represented by the BDDs $\Delta_1, \ldots, \Delta_{|\Z|}$ and $\Phi$ can then be given directly to the symbolic game-solving algorithm implemented in {\Syft} to compute a strategy.

\subsection{Belief-State Construction} \label{sec:belief-states}

The belief-states approach described in~\cite{GiacomoV16} is based on a standard construction used in planning under partial observability~\cite{GoldmanB96,BonetG00,BryceKS06,MaliahBKS14}. Given a DFA $D$ for the {\ltlf} formula $\varphi$, this approach constructs a new DFA $B$ where the state space is formed of \emph{belief states}, which are sets of states of $D$ representing the possible states in which the game can be given the information observed by the system. Since $B$ is exponential in $D$, and $D$ is in the worst case doubly-exponential in $\varphi$, in the worst case this approach is triple-exponential.

As pointed out in~\cite{GiacomoV16}, the belief-state construction is equivalent to starting the projection-based construction outlined in Section~\ref{sec:double-negation} from a DFA $D$ (constructed normally by {\Mona}) rather than an {\ltlf} formula. In this case, rather than negating the formula, we simply complement $D$. Since a DFA is a special case of an NFA, the last three steps can be performed exactly in the same way as in the projection-based approach. Therefore, the belief-state construction can likewise be performed symbolically, potentially saving one exponential as well. Note that the subset construction used to determinize the NFA now constructs the belief states. The existential quantification in the definition of $\Delta_j$ can be interpreted as 
adding to the belief state every state $s_j$ for which there is a possibility of the unobservable inputs having moved the automaton to $s_j$.
Finally, note that since the set $F$ of accepting states of $D$ was complemented in the first step, the final BDD for the accepting states of $A$ is $\Phi = \neg \bigvee_{s_i \not\in F} z_i = \bigwedge_{s_i \not\in F} \neg z_i$. This corresponds to the accepting belief-states being those that contain only accepting states of $D$, i.e., only those where the system can be sure that it is in an accepting state.

The fact that the DFA for an {\ltlf} formula may be doubly-exponential, while a NFA is at most exponential, seems to reinforce the notion that the projection-based approach is strictly better. In practice, however, it has been observed that fully-minimized DFA (as is the case of the DFAs produced by {\Mona}) are rarely doubly-exponential, and in some cases when NFA are determinized and minimized they actually become smaller~\cite{TabakovRV12}. Therefore, it is important to compare the two approaches empirically as well, which we do in Section~\ref{sec:experiments}.

\subsection{MSO Construction} \label{sec:mso}

Although the above two approaches were the only ones presented in~\cite{GiacomoV16}, the synthesis framework employed by {\Syft} naturally suggests a third approach for synthesis under partial observability. In the second step of {\Syft}'s workflow, {\Mona} is used to convert the first-order-logic formula $fol(\varphi)$ into a DFA. {\Mona}, however, can handle not only first-order formulas, but also more general formulas in \emph{monadic second-order logic} (MSO)~\cite{Buchi1960}. MSO can easily model quantification over traces, allowing us to express in MSO the language of traces over $Obs \cup \Y$ such that for all traces over $Unobs$ the {\ltlf} formula $\varphi$ is satisfied. This language is represented simply by the formula $\forall U_1 . \ldots \forall U_n . fol(\varphi)$, where each $U_i$ is a second-order variable corresponding to one of the unobservable propositions. Thus, by simply adding to {\Syft}'s workflow the step of quantifying the unobservable propositions, we can get {\Syft} to solve the synthesis problem under partial observability. The following theorem follows directly from the MSO semantics and states the correctness of this approach.

\begin{theorem}
A strategy for the DFA game specified by the MSO formula $\forall U_1 . \ldots \forall U_n . fol(\varphi)$ is winning for the system iff that strategy is a solution to the synthesis problem for $\varphi$ under partial observability.
\end{theorem}

Interestingly, the procedure used by {\Mona} to construct the DFA for this MSO formula resembles the projection-based construction. {\Mona} uses a syntactic approach for constructing DFAs, first rewriting $\forall U_1 . \ldots \forall U_n . fol(\varphi)$ as $\neg (\exists U_1 . \ldots \exists U_n . \neg fol(\varphi))$, then building a DFA for $fol(\varphi)$ and applying complementation and projection as appropriate. Thus, it follows the same sequence of steps outlined in Section~\ref{sec:double-negation}. Note, however, that {\Mona} not only starts with a DFA, like in the belief-states construction, but also determinizes the intermediate automata after every projection operation, since it does not have an internal representation for NFAs. This means that although the final DFA is minimal, and therefore doubly-exponential at most, it is possible that the subset-construction operation may add a third exponential to the running time. On the other hand, {\Mona} minimizes intermediate DFAs after every operation, which can actually make them significantly smaller and may improve the running time in practice.
Furthermore, because the final DFA output by {\Mona} is fully minimized, the number of states may be much smaller than that of the final DFAs produced by the other procedures, for which subset construction is performed symbolically and therefore does not go through minimization. On the other hand, because the final DFA is not generated directly in symbolic representation, if the number of states is large the DFA construction is more likely to fail. Considering these points, it is not clear in general how this approach would compare with the others, and answering this question requires an experimental evaluation.
\section{Experimental Evaluation} \label{sec:experiments}

As mentioned in the previous section, theory is not necessarily a good indicator for performance in practice. There are a number of factors that are not factored into the theoretical analysis but can affect the performance of the three approaches described in Section~\ref{sec:approaches}, including the difference in practice of DFA vs. NFA size, the DFA-construction algorithm implemented by {\Mona}, and the use of symbolic representation. Therefore, it is essential to complement the theory with an empirical evaluation in order to determine the relative advantage of each of the three approaches. We first present three families of benchmarks that we used in our evaluation, and then describe our experimental setup and results.

\subsection{Benchmarks}

To model settings where the system must behave strategically in the presence of partial observability, keeping track of information learned during interaction with the environment, we constructed {\ltlf} specifications describing games with incomplete information. We constructed three benchmark families for our evaluation. Below we provide a brief high-level description of each family, then present and explain the general form of the {\ltlf} specification in each case, indicating the observable inputs, unobservable inputs and outputs, as well as whether the specification is realizable or not and some intuition about the winning strategy if it is realizable. 
It is worth noting that the first two benchmark families are simpler in the sense that they use only the $X$, $G$ and $F$ operators, while the third family also uses the more general $U$ operator. Nevertheless, we obtain the same general conclusions from all three of them.

\subsubsection{Moving-Target}

The environment controls a target moving along a line with $n$ positions. The target's location and movement are unknown to the system, who at every turn tries to guess where the target is.
\begin{align*}
    &\varphi_{Target} = G(\texttt{exactly-one}(target_1, \ldots, target_n)) \\
    &\varphi_{Move} = G (X\,true \rightarrow \texttt{move-left-or-right}(target_1, \ldots, target_n)) \\
    &\varphi_{Hit} = \bigwedge^n_{i = 1} G((target_i \land guess_i) \rightarrow hit) \qquad \varphi_{System} = G(\texttt{exactly-one}(guess_1, \ldots, guess_n)) \land F hit \\
    &\textbf{Full specification: } (\varphi_{Target} \land \varphi_{Move} \land \varphi_{Hit}) \rightarrow \varphi_{System}
\end{align*}

$target_1, \ldots, target_n$ are unobservable input variables such that $target_i$ is true if the target is in position $i$ of the line. Exactly one $target_i$ variable must be true at a given time (as specified in $\varphi_{Target}$). We omit details of the subformula $\texttt{move-left-or-right}(target_1, \ldots, target_n)$ in $\varphi_{Move}$, but it suffices to know that it establishes a relation between the values of $target_1, \ldots, target_n$ in adjacent time steps, namely that the target must always move to the position immediately to the left or to the right of the previous position (and cannot stay in the same position). 
If the target is in position $1$ or $n$, then the only option is for it to move to position $2$ or $n-1$, respectively.
$guess_1, \ldots, guess_n$ are output variables such that $guess_i$ is set to true to guess that the target is in position $i$. $hit$ is an observable input variable that is set to true if the guess is correct (as specified in $\varphi_{Hit}$). The system can only make one guess at a time, and it wins if it guesses correctly ($\varphi_{System}$). All specifications in this family are realizable regardless of the value of $n$. The winning strategy for the system player is based on two rules: first, if the target is not in position $2$ at time $t$, then it cannot be in position $1$ at time $t + 1$ (same for $n - 1$ and $n$); second, if the target is neither in position $i - 1$ nor $i + 1$ at time $t$, then it cannot be in position $i$ at time $t + 1$. Using these two rules, the system can guess in such a way that it narrows down the positions the target can be in over time, guaranteeing that it will hit the target eventually.

\subsubsection{Coin-Game}

This is an $n$-coin generalization of the game described in~\cite{DoyenR11}. Every turn the system chooses a coin to flip, and wins once all coins are heads. The environment reports whether the coin was flipped to heads or tails, and can secretly swap the two coins adjacent to it.
\begin{align*}
    &\varphi_{Init} = \texttt{exactly-one}(\neg coin_1, \ldots, \neg coin_n) \qquad \varphi_{Valid} = X\,G\,(valid \leftrightarrow \texttt{exactly-one}(flip_1, \ldots, flip_n)) \\
    &\varphi_{Update} = \bigwedge^n_{i = 1} G(X\,valid \rightarrow \texttt{update}_i(flip_1, \ldots, flip_n, coin_1, \ldots, coin_n, swap)) \\
    &\varphi_{Heads} = X\,G\,(valid \rightarrow (heads \leftrightarrow \bigvee^n_{i = 1} (flip_i \land coin_i))) \qquad \varphi_{System} = F \bigwedge^n_{i = 0} coin_i \\
    &\textbf{Full specification: } (\varphi_{Init} \land \varphi_{Valid} \land \varphi_{Heads} \land \varphi_{Update}) \rightarrow \varphi_{System}
\end{align*}

$coin_1, \ldots, coin_n$ are unobservable input variables such that $coin_i$ is true if the $i$-th coin is heads. Initially, exactly one coin is heads ($\varphi_{Init}$). $flip_1, \ldots, flip_n$ are output variables such that $flip_i$ is set to true to flip the $i$-th coin.
The system may only flip a single coin, otherwise the observable input variable $valid$ is set to false ($\varphi_{Valid}$).
The unobservable input variable $swap$ indicates whether the environment decides to swap the two coins adjacent to the coin flipped by the system. If the move is valid, the state of the coins is updated by the environment ($\varphi_{Update}$). The subformula $\texttt{update}_i(flip_1, \ldots, flip_n, coin_1, \ldots, coin_n, swap)$ in $\varphi_{Update}$ expresses how $coin_i$ variables are updated.
We omit details, but intuitively when $flip_i$ is true $coin_i$ changes value, and if additionally $swap$ is true then $coin_{(i - 1) \mod n}$ and $coin_{(i + 1) \mod n}$ have their values swapped. If the move is valid the environment also reports using the observable input variable $heads$ whether the flipped coin was flipped to heads or not ($\varphi_{Heads}$). The system wins if all coins are flipped to heads ($\varphi_{System}$). The specification is unrealizable for $n = 3$ and realizable for all $n > 3$. The winning strategy for even $n$ is simple: flip all even-numbered coins to heads, then flip all odd-numbered coins to heads. This prevents the environment from secretly swapping a flipped coin with an unflipped coin, since it is only able to swap coins that are adjacent to the last coin that was flipped. A similar strategy also works with some adjustment for odd $n$, except for $n = 3$, where flipping a coin always gives the environment the opportunity to swap the other two.

\subsubsection{Private-Peek}

This family of benchmarks is based on the game described in~\cite{Reif84}, which is an incomplete-information version of the \emph{Peek} game from~\cite{StockmeyerC79}.
Players push plates with holes in and out of a box. Depending on their configuration they might uncover holes on the box so that one or the other player can peek through to the other side. The first player able to do so wins. Each player has $n$ plates to control and $m$ holes that they might be able to peek through, and only half of the plates (rounded up) of the environment are visible to the system. As the positions of the holes in each plate are arbitrary, we can generate multiple instances for each value of $m$ and $n$ by randomly selecting the hole positions.
\begin{align*}
    &\varphi^p_{In} = \bigwedge^n_{i=1} plate^p_i \qquad \varphi^p_{Wait} = \bigwedge^n_{i=1} G(X \neg turn^p \rightarrow (X\,plate^p_i \leftrightarrow plate^p_i)) \\
    &\varphi^p_{Move} = \bigwedge^n_{i=1} G(X\,turn^p \rightarrow \texttt{at-most-one}(X\,plate^p_1 \leftrightarrow \neg plate^p_i, \ldots, X\,plate^p_n \leftrightarrow \neg plate^p_n)) \\
    &\varphi^p_{Peek} = \bigwedge^m_{j = 1} G(peek^p_j \leftrightarrow \texttt{random-cube}^p_j(plate^e_1, \ldots, plate^e_n, plate^s_1, \ldots, plate^s_n)) \\
    &\varphi_{Turn} = \neg turn^e \land \neg turn^s \land X\,turn^s \land X\,G(turn^s \leftrightarrow \neg turn^e) \land X\,G(X\,true \rightarrow (X\,turn^s \leftrightarrow turn^e)) \\
    &\varphi_{Goal} = \left(\bigwedge^m_{j = 1} (turn^e \rightarrow \neg peek^e_j)\right) U \left(turn^s \land \bigvee^m_{j = 1} peek^s_j\right) \\
    &\textbf{Full specification: } (\varphi^e_{In} \land \varphi^e_{Wait} \land \varphi^e_{Move} \land \varphi^e_{Peek} \land \varphi^s_{Peek}) \rightarrow (\varphi_{Turn} \land \varphi^s_{In} \land \varphi^s_{Wait} \land \varphi^s_{Move} \land \varphi_{Goal})
\end{align*}

The game alternates between system and environment turns. Output variables $turn^s$ and $turn^e$ are used to keep track of turns, and are set to true when it is the system's and the environment's turn, respectively. In the first timestep, which serves just to set up the initial state of the game, both $turn^s$ and $turn^e$ are set to false. The first turn of the system occurs in the second timestep, and turns alternate after that ($\varphi_{Turn}$). Variables $plate^p_1, \ldots, plate^p_n$ are input variables for $p = e$ and output variables for $p = s$, and $plate^p_i$ is true if the $i$-th plate of player $p$ is in, and false if it is out. Initially, all plates are in ($\varphi^p_{In}$), and on their own turn each player can choose to slide at most one of them in or out ($\varphi^p_{Move}$). On the opponent's turn the player cannot move their plates ($\varphi^p_{Wait}$).

$peek^p_1, \ldots, peek^p_m$ are input variables such that $peek^p_j$ is true if player $p \in \{e, s\}$ can peek through their $j$-th hole to the other side. The system wins if it is able to peek through one of its holes before the environment can ($\varphi^{Goal}$). The set of configurations that determines whether the $j$-th hole of player $p \in \{e, s\}$ is uncovered is encoded in $\varphi^p_{Peek}$ by the formula $\texttt{random-cube}^p_j(plate^e_1, \ldots, plate^e_n, plate^s_1, \ldots, plate^s_n)$, which is generated randomly by selecting a subset of the variables $plate^e_1, \ldots, plate^e_n, plate^s_1, \ldots, plate^s_n$, each in either positive of negative form, and taking their conjunction. This encoding is justified by the equivalence of \emph{Peek} to a formula game played over a formula in disjunctive normal form~\cite{StockmeyerC79, Reif84}. Intuitively, if $plate^p_i$ appears in positive (respectively, negative) form it means that the $i$-th plate of player $p$ needs to be in (out) to uncover the hole. If $plate^p_i$ does not appear at all, then either configuration works. In this way, we can generate multiple random instances of the \emph{Private-Peek} family for each $m$ and $n$. For our experiments, each variable has a $1/2$ chance of being selected for a given random cube, and each selected variable is negated with probability $1/2$ as well. Whether the instance is realizable or unrealizable depends on the random formulas generated. Out of the input variables, $peek^e_1, \ldots, peek^e_m, plate^e_1, \ldots, plate^e_{\lceil \frac{n}{2} \rceil}$ are unobservable. This corresponds to barriers keeping the system from seeing the holes on the environment side, as well as half of the environment plates (rounded up).

\subsection{Experimental Setup and Results}

We generated instances with $n$ varying from $2$ to $10$ for the \emph{Moving-Target} benchmarks and $3$ to $10$ for the \emph{Coin-Game} benchmarks. For the \emph{Private-Peek} benchmarks, we varied $n$ and $m$ from $1$ to $4$ and generated $30$ random instances for each combination using the procedure described 
above. We report the median results for each combination of $n$ and $m$.
We ran all experiments on a single node of a high-performance cluster consisting of an Intel Xeon processor runninng at 2.6~GHz. Experiments had 32~GB of memory available and a timeout of 8 hours. Failed instances are due to either timeouts or memouts.

The {\ltlf}-synthesis procedure implemented in {\Syft} consists of two phases, one explicit and one symbolic. The first phase is the construction of an explicit automaton by {\Mona}, and the second phase is the conversion of this automaton to a symbolic representation followed by the symbolic fixpoint computation used to compute the winning strategy. We analyze how the three approaches perform in each of these two phases, and then see how they contribute to the overall performance. Although {\Syft} uses a fixed variable ordering for BDDs, in order to reduce the impact that a single variable ordering has on BDD sizes and make for a more fair comparison between the different approaches, we enabled dynamic variable reordering~\cite{FeltYBS93}, which tries to optimize the ordering of variables on the fly during execution.

\subsubsection{Explicit Phase}

\begin{figure}
     \centering
     \begin{subfigure}[b]{0.49\textwidth}
         \centering
         \includegraphics[width=\textwidth]{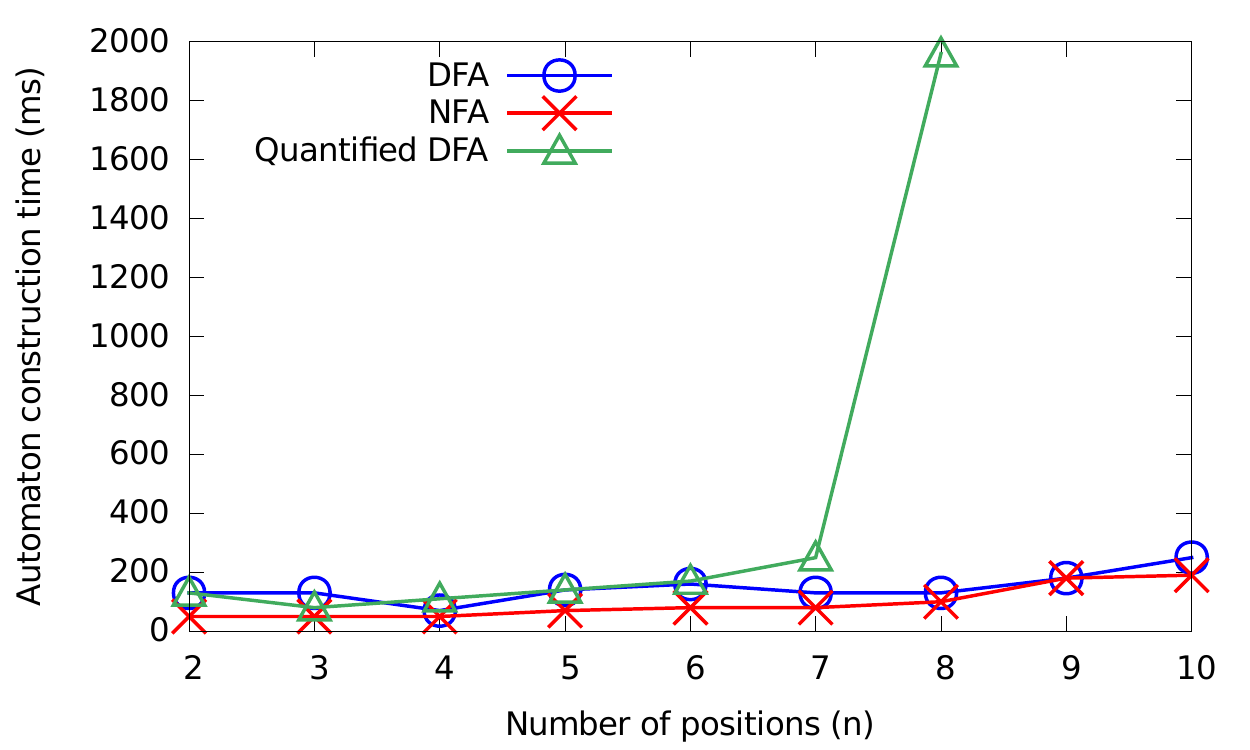}
         \caption{Automaton construction time (\emph{Moving-Target})}
         \label{fig:target-mona}
     \end{subfigure}
     \hfill
     \begin{subfigure}[b]{0.49\textwidth}
         \centering
         \includegraphics[width=\textwidth]{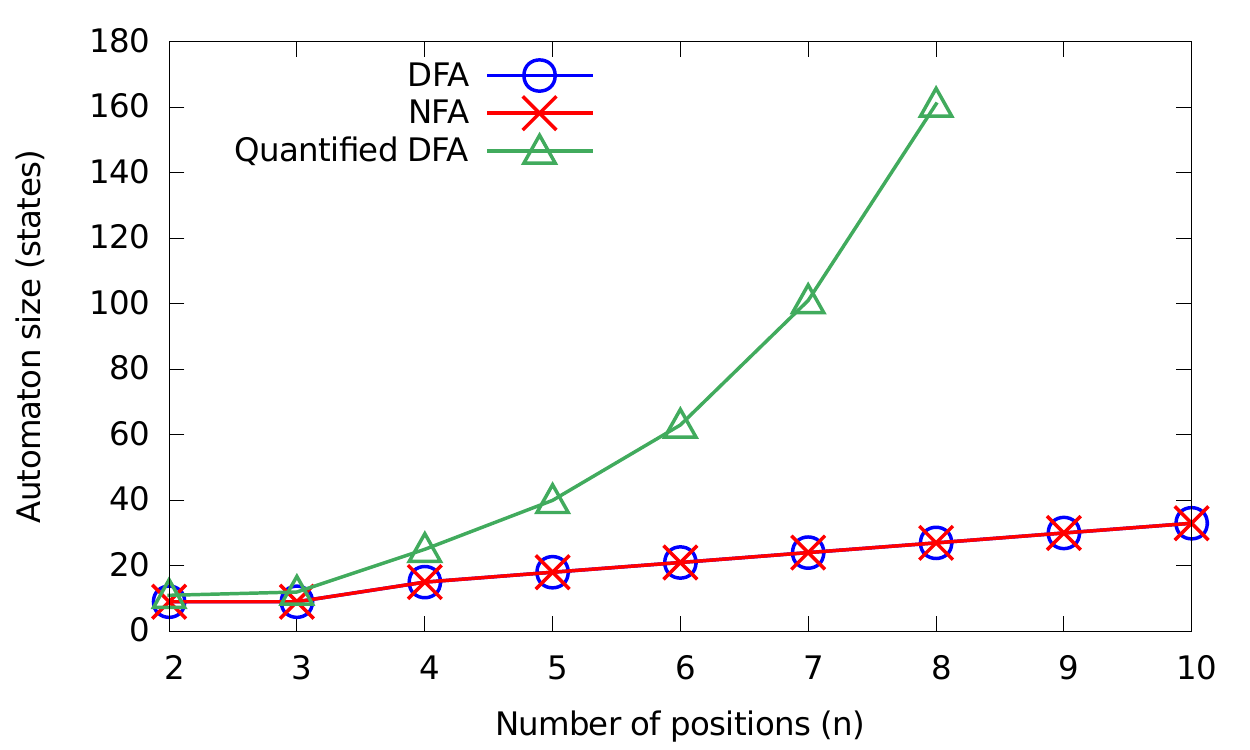}
         \caption{Number of automaton states (\emph{Moving-Target})}
         \label{fig:target-size}
     \end{subfigure}
     
     \bigskip
     
     \begin{subfigure}[b]{0.49\textwidth}
         \centering
         \includegraphics[width=\textwidth]{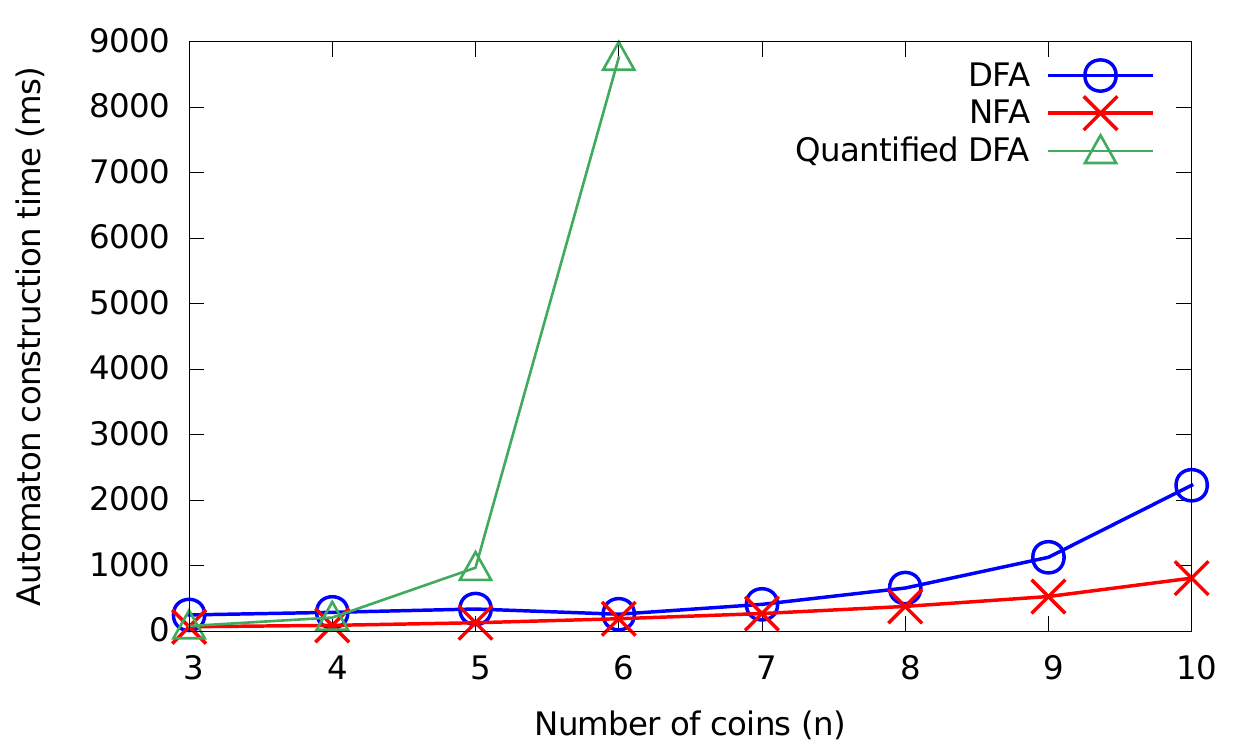}
         \caption{Automaton construction time (\emph{Coin-Game})}
         \label{fig:coins-mona}
     \end{subfigure}
     \hfill
     \begin{subfigure}[b]{0.49\textwidth}
         \centering
         \includegraphics[width=\textwidth]{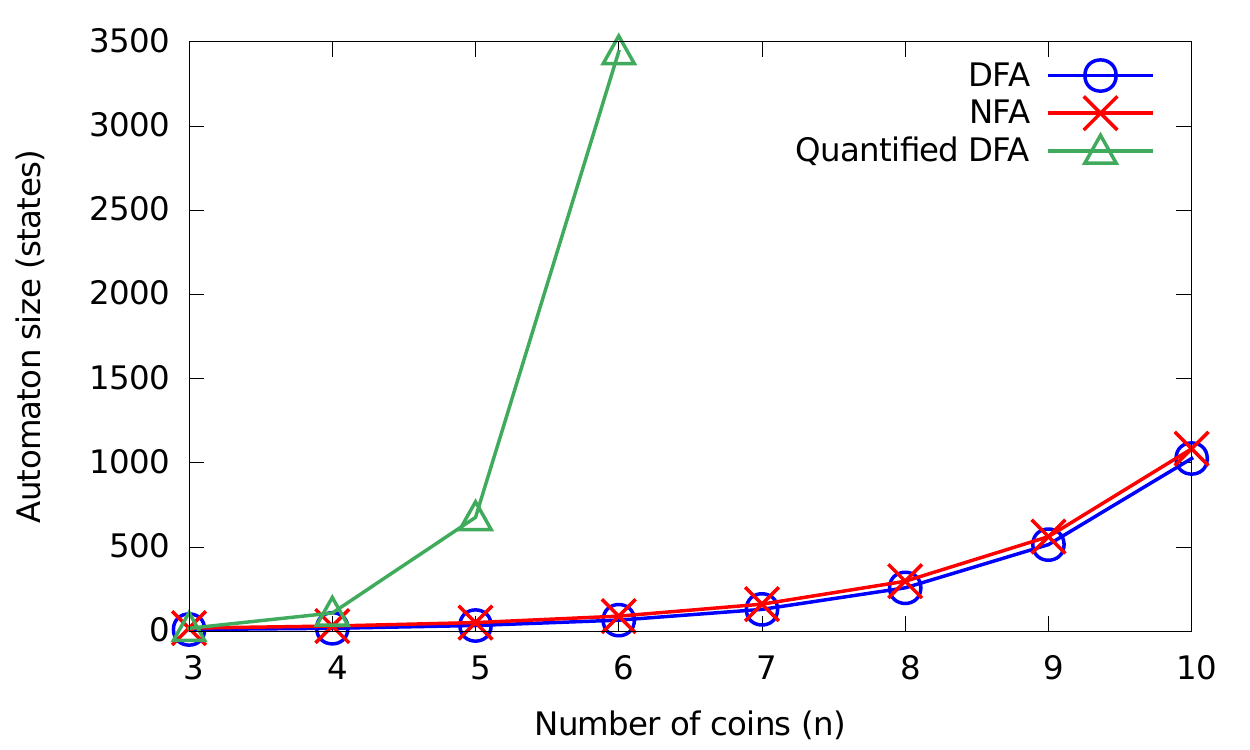}
         \caption{Number of automaton states (\emph{Coin-Game})}
         \label{fig:coins-size}
     \end{subfigure}
     
    %\bigskip
    \medskip
     
     \begin{subfigure}[b]{0.49\textwidth}
         \centering
         \includegraphics[width=\textwidth]{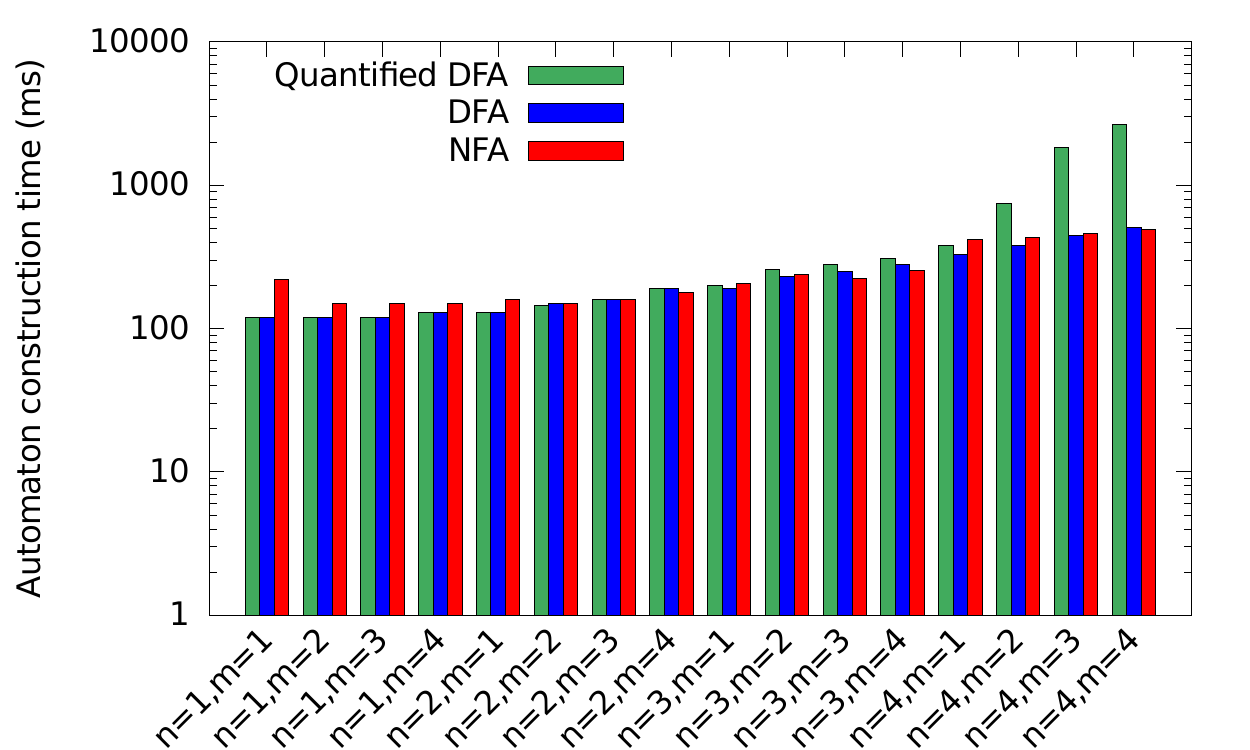}
         \caption{Automaton construction time (\emph{Private-Peek})}
         \label{fig:peek-mona}
     \end{subfigure}
     \hfill
     \begin{subfigure}[b]{0.49\textwidth}
         \centering
         \includegraphics[width=\textwidth]{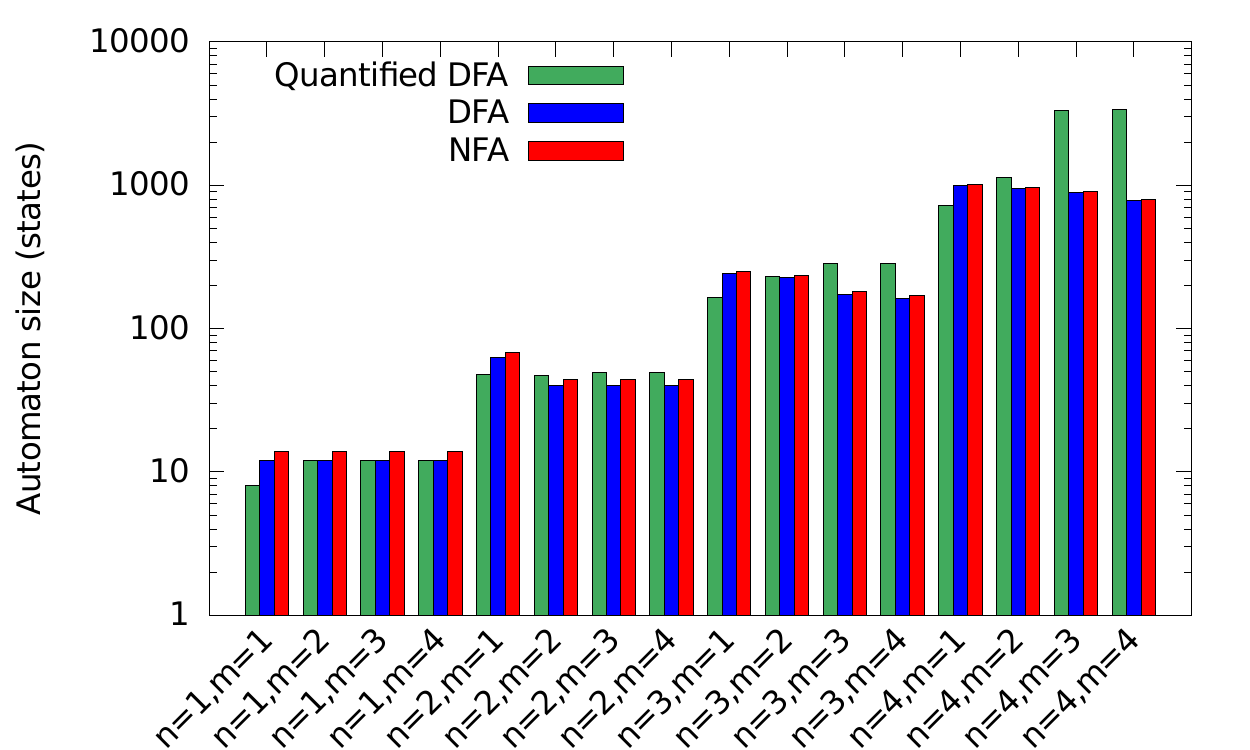}
         \caption{Number of automaton states (\emph{Private-Peek})}
         \label{fig:peek-size}
     \end{subfigure}
        \caption{Automaton construction times and number of automaton states for different values of $n$ and $m$ for each benchmark family. Values for the \emph{Private-Peek} benchmarks are the median of the $30$ random instances, and are presented in log scale.}
        \label{fig:mona}
\end{figure}

We first analyze the explicit
phase across the three approaches. Recall that {\Mona} constructs an explicit NFA in the projection-based approach (via a DFA for the reverse language) and explicit DFAs for the other two approaches (in the MSO approach, with the unobservable inputs universally quantified). Recall that the NFA can be at most exponential in the size of the original {\ltlf} formula, while the DFA (quantified or not) will be doubly-exponential in the worst case.
Interestingly, as can be seen in  Figures~\ref{fig:target-mona},~\ref{fig:coins-mona} and~\ref{fig:peek-mona}, there is not a big difference in running time between constructing a DFA and an NFA for the {\ltlf} formula. In fact, as Figures~\ref{fig:target-size},~\ref{fig:coins-size} and~\ref{fig:peek-size} show, for the \emph{Moving-Target} formulas the DFA and NFA of the formula have exactly the same number of states, while for the \emph{Coin-Game} and \emph{Private-Peek} benchmarks the NFA was in fact slightly larger. This happens in part because moves in these games are reversible, meaning that the DFA for the reverse language (which is reversed to produce the NFA) has a similar structure to the original language. Nevertheless, these results reinforce the observation from~\cite{TabakovRV12} that the exponential gap between DFA and NFA often does not occur in practice when the DFA is minimized. As this gap was the central reason for the exponential gap in complexity between the belief-states and projection-based constructions, this result is significant to highlight how theoretical analysis may not always accurately predict behavior in practice.
On the other hand, the universally-quantified DFA, despite having the same worst-case as the non-quantified DFA, in practice grows much faster. In the \emph{Private-Peek} instances it tends to be smaller than the DFA for low values of $m$ but quickly grow to surpass it as $m$ increases. In the \emph{Moving-Target} and \emph{Coins-Game} families the construction  of the quantified DFA also could not be completed for larger values of $n$ ($> 8$ for \emph{Moving-Target} and $> 6$ for \emph{Coins-Game}), not only due to the size of the final DFA but likely also due to the overhead of {\Mona}'s construction algorithm, as mentioned in Section~\ref{sec:mso}.

\subsubsection{Symbolic Phase}

\begin{figure}[t]
     \centering
     \begin{subfigure}[b]{0.49\textwidth}
         \centering
         \includegraphics[width=\textwidth]{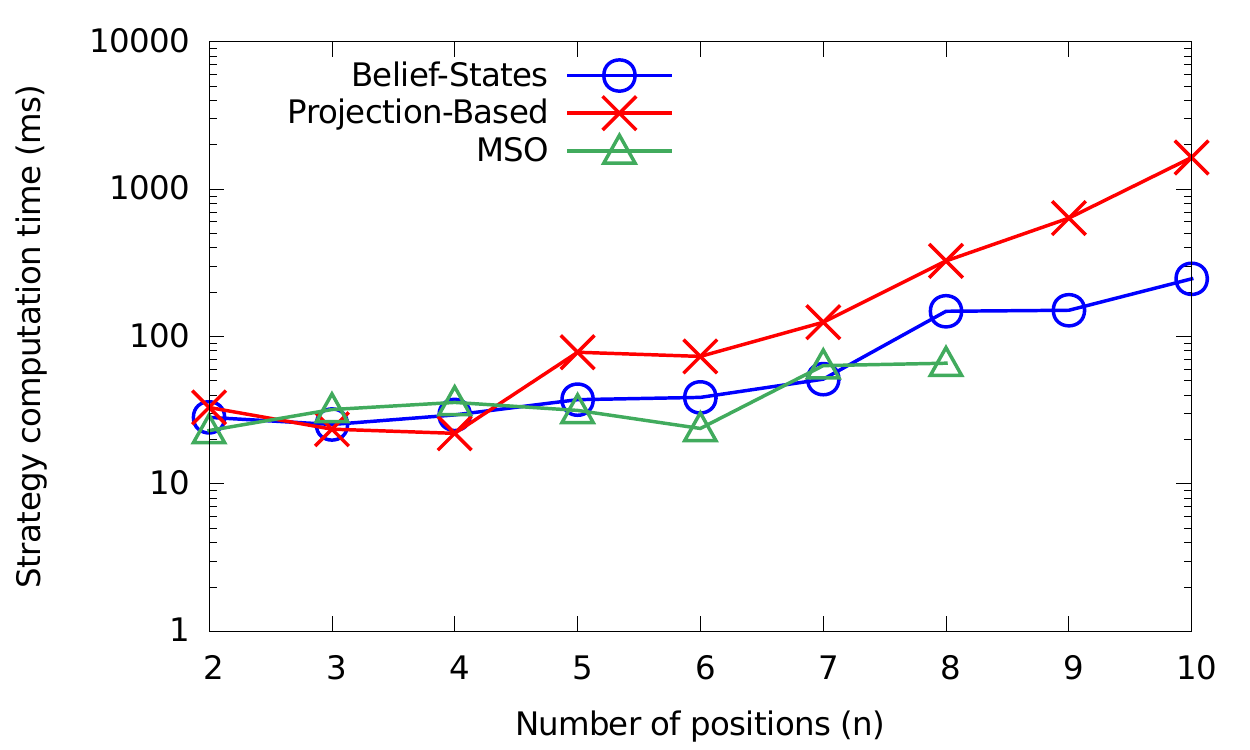}
         \caption{\emph{Moving-Target}}
         \label{fig:target-syft}
     \end{subfigure}
     \hfill
     \begin{subfigure}[b]{0.49\textwidth}
         \centering
         \includegraphics[width=\textwidth]{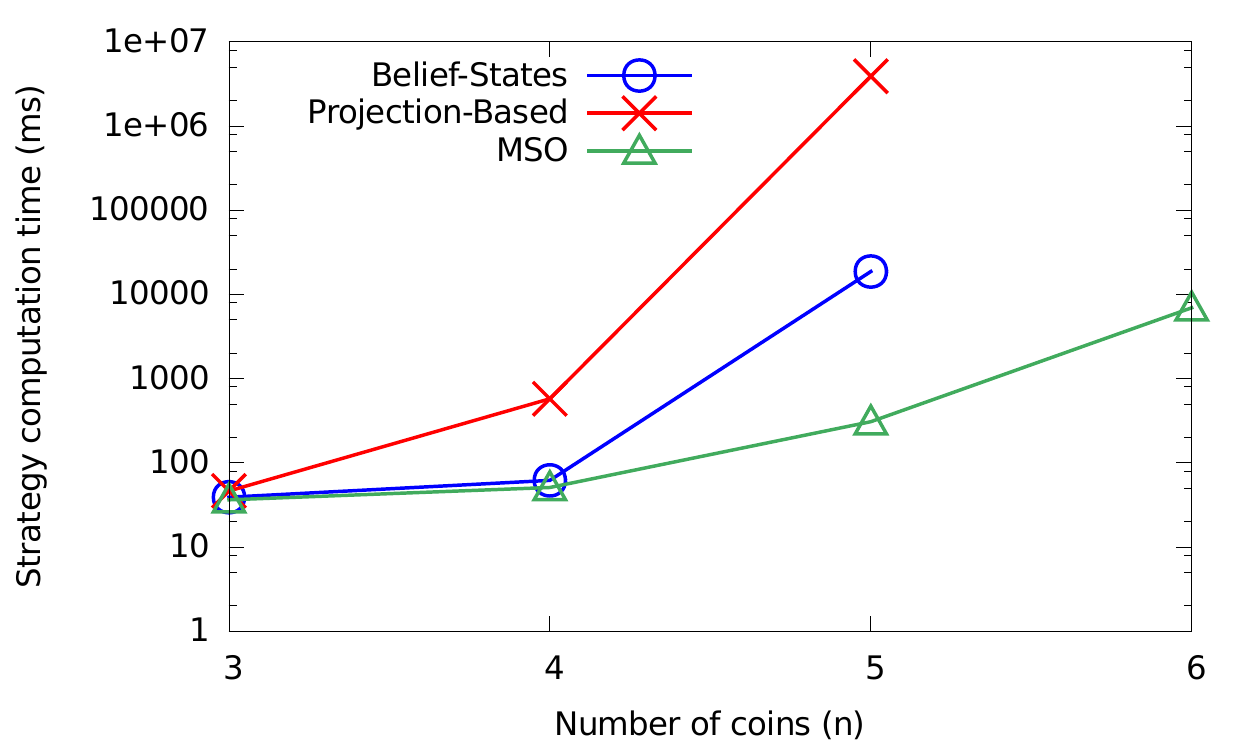}
         \caption{\emph{Coin-Game}}
         \label{fig:coins-syft}
     \end{subfigure}
     
     \medskip
     
     \begin{subfigure}[b]{0.49\textwidth}
         \centering
         \includegraphics[width=\textwidth]{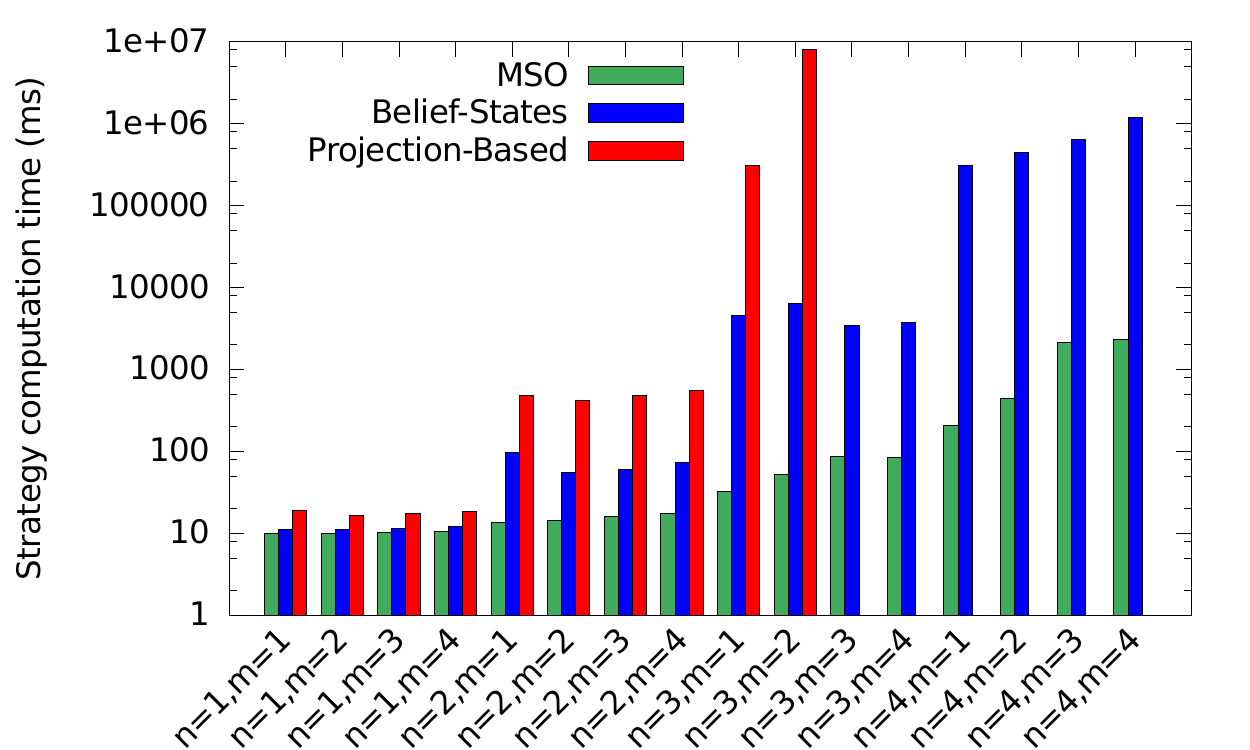}
         \caption{\emph{Private-Peek}}
         \label{fig:peek-syft}
     \end{subfigure}
        \caption{Time taken to solve the DFA game for each value of $n$ and $m$ for each benchmark family, in log scale. Values for the \emph{Private-Peek} family are the median of the 30 random instances, and missing values mean the median could not be computed because most random instances timed out.}
        \label{fig:syft}
\end{figure}

In the second phase, the explicit automata generated by {\Mona} are converted into symbolic DFAs. In the case of the projection-based and belief-states constructions, this means performing symbolic subset construction as described in Sections~\ref{sec:double-negation} and~\ref{sec:belief-states}, which means that the number of state variables in the symbolic representation is equal to the number of states in the explicit automaton. As each assignment of the state variables represents a state, this represents an exponential blowup in the state space caused by subset construction. On the other hand, in the MSO approach no subset construction is necessary and the state space of the universally-quantified DFA produced by {\Mona} can be encoded in a logarithmic number of state variables. As a result, even though the quantified DFA is significantly larger, the number of state variables used in the symbolic representation of the MSO approach is smaller than in the other approaches. Even so, the size of the initial symbolic representation, measured in total number of nodes in the BDDs for the transition relation and accepting states, is larger for the MSO approach. This might be because the symbolic representations generated by subset construction display more structure than the one generated by a logarithmic encoding.

As can be seen in Figure~\ref{fig:syft}, however, when computing the winning strategy from the symbolic representation of the DFA game the MSO approach was the fastest across the board. This suggests that the size of the implicit state space, represented by the number of state variables, is a more important factor in the performance of the second phase than the initial symbolic representation of the automaton.
Note that the lack of results for larger $n$ for the MSO approach in the \emph{Moving-Target} and \emph{Coin-Game} benchmarks is not due to the performance of the algorithm in this phase, but rather due to the quantified DFA not being able to be constructed in the first phase. Interestingly, even though the other two approaches were able to construct automata up to $n = 10$ for the \emph{Coin-Game} benchmarks, they failed to solve the game for $n > 5$, while the MSO approach can still construct the automaton and solve the game for $n = 6$. Perhaps surprisingly, the projection-based approach had the worst performance overall. In addition to the number of states of the NFA being equal or slightly larger than the DFA, the symbolic representation and strategy-computation time were also significantly larger. It was also unable to solve the game in most cases for larger values of $n$ and $m$ in the \emph{Private-Peek} family. This serves as final evidence of the importance of complementing theoretical analysis with empirical evaluation.

\subsubsection{End-to-end Picture}

Given the results presented in the two phases above, two major conclusions can be drawn. The first is that when comparing between the MSO approach and the other two, there is a tradeoff between the explicit and symbolic phases. The MSO approach produces a much larger explicit automaton and requires more time to do so, and as a result is more likely to run out of time or memory to complete the first phase. On the other hand, the fact that this automaton is minimized, compared to the symbolic DFA generated by the other two approaches, leads to a significantly better performance when solving the game in the second phase. If we add the times for the two phases, the MSO approach takes slightly longer for the \emph{Moving-Target} benchmarks, but performs significantly better in the \emph{Coin-Game} and \emph{Private-Peek} benchmarks. Overall, the MSO approach seems to be a good option as long as the construction of the quantified DFA can be completed. The second conclusion is that, unlike what the theory seems to suggest, the performance of the projection-based approach is more often than not worse than that of the belief-states construction. Although in the worst case there might be an exponential gap between NFA and DFA, this behavior has not been observed in practice, which closes the gap between the 2EXPTIME and 3EXPTIME complexity of the two constructions. Furthermore, in the second phase the projection-based approach produced a less efficient symbolic representation, and ultimately required more time to solve the game. These results also highlight the necessity of considering algorithmic details that are hard to account for in a purely theoretical analysis, such as the use of symbolic techniques.

\section{Discussion}

We have undertaken the first steps to
bring synthesis under partial observability,
which previously inhabited only the realm of pure theory,
closer to practical application. Much work still remains to be done to scale to real-world scenarios, but supported by the conceptual simplicity of {\ltlf} synthesis compared to {\ltl} synthesis and the availability of efficient tools such as {\Mona} and {\Syft} we have presented the first implementation and empirical evaluation of algorithms for temporal synthesis under partial observability. Our experimental evaluation showed that the choice of algorithm for {\ltlf} synthesis under partial observability is not as straightforward in practice as the theoretical analysis from~\cite{GiacomoV16} would suggest.

First, although the projection-based approach is exponentially better than the belief-states construction in theory, this advantage depends on the assumption that the NFA constructed is smaller than the corresponding DFA. In our examples this assumption was violated, negating the advantage of this approach. Second, the use of symbolic algorithms means that synthesis performance depends not only on the number of automaton states, but also on the size of its symbolic representation. The projection-based approach's giving rise to a less efficient symbolic representation had a larger effect in the results than automata size. Finally, {\Mona} enabled the introduction of an MSO-based approach that has its own pros and cons. Although that approach can be more efficient for computing a winning strategy, it pays a price during explicit DFA construction and may fail during this phase if the DFA is large. It would be interesting to investigate how the improvement of explicit DFA construction algorithms could help the MSO approach overcome this obstacle, and how this would change the general picture. As for the other algorithms, the priority should be to bridge their gap in performance during the symbolic fixpoint computation, which as observed in Section~\ref{sec:experiments} is primarily due to the lack of minimization of the final symbolic DFA. This is not a trivial problem to solve, as minimization of symbolic state spaces is not always effective~\cite{FislerV02}.

Our observations suggest that rather than there being a single best algorithm for {\ltlf} synthesis under partial observability, we need a portfolio of algorithms, and the best option will likely depend on the nature of the problem being solved. When the quantified DFA can be constructed explicitly within the available time and memory, the MSO approach will probably excel. For most other cases, the belief-states construction is likely to be the best option, the exception being extreme cases where the DFA is doubly exponential, and therefore the use of the NFA by the projection-based approach provides an advantage. This is a very different result than what is suggested by the purely theoretical analysis, showing the importance of studying the problem empirically as well.

\section*{Acknowledgments}

Work supported in part by NSF grants IIS-1527668, CCF-1704883,
IIS-1830549, and an award from the Maryland Procurement Office.

%\nocite{*}
\bibliographystyle{eptcs}
\bibliography{gandalf}

\begin{thebibliography}{10}
\providecommand{\bibitemdeclare}[2]{}
\providecommand{\surnamestart}{}
\providecommand{\surnameend}{}
\providecommand{\urlprefix}{Available at }
\providecommand{\url}[1]{\texttt{#1}}
\providecommand{\href}[2]{\texttt{#2}}
\providecommand{\urlalt}[2]{\href{#1}{#2}}
\providecommand{\doi}[1]{doi:\urlalt{http://dx.doi.org/#1}{#1}}
\providecommand{\bibinfo}[2]{#2}

\bibitemdeclare{inproceedings}{BohyBFJR12}
\bibitem{BohyBFJR12}
\bibinfo{author}{Aaron \surnamestart Bohy\surnameend},
  \bibinfo{author}{V{\'{e}}ronique \surnamestart Bruy{\`{e}}re\surnameend},
  \bibinfo{author}{Emmanuel \surnamestart Filiot\surnameend},
  \bibinfo{author}{Naiyong \surnamestart Jin\surnameend} \&
  \bibinfo{author}{Jean{-}Fran{\c{c}}ois \surnamestart Raskin\surnameend}
  (\bibinfo{year}{2012}): \emph{\bibinfo{title}{{Acacia+, a Tool for {LTL}
  Synthesis}}}.
\newblock In: {\sl \bibinfo{booktitle}{{CAV}}}, pp. \bibinfo{pages}{652--657},
  \doi{10.1007/978-3-642-31424-7\_45}.

\bibitemdeclare{inproceedings}{BonetG00}
\bibitem{BonetG00}
\bibinfo{author}{Blai \surnamestart Bonet\surnameend} \&
  \bibinfo{author}{Hector \surnamestart Geffner\surnameend}
  (\bibinfo{year}{2000}): \emph{\bibinfo{title}{{Planning with Incomplete
  Information as Heuristic Search in Belief Space}}}.
\newblock In: {\sl \bibinfo{booktitle}{Proceedings of the Fifth International
  Conference on Artificial Intelligence Planning Systems}}, pp.
  \bibinfo{pages}{52--61}.

\bibitemdeclare{article}{Bryant86}
\bibitem{Bryant86}
\bibinfo{author}{Randal~E. \surnamestart Bryant\surnameend}
  (\bibinfo{year}{1986}): \emph{\bibinfo{title}{{Graph-Based Algorithms for
  Boolean Function Manipulation}}}.
\newblock {\sl \bibinfo{journal}{{IEEE} Trans. Computers}}
  \bibinfo{volume}{35}(\bibinfo{number}{8}), pp. \bibinfo{pages}{677--691},
  \doi{10.1109/TC.1986.1676819}.

\bibitemdeclare{article}{BryceKS06}
\bibitem{BryceKS06}
\bibinfo{author}{Daniel \surnamestart Bryce\surnameend},
  \bibinfo{author}{Subbarao \surnamestart Kambhampati\surnameend} \&
  \bibinfo{author}{David~E. \surnamestart Smith\surnameend}
  (\bibinfo{year}{2006}): \emph{\bibinfo{title}{{Planning Graph Heuristics for
  Belief Space Search}}}.
\newblock {\sl \bibinfo{journal}{J. Artif. Intell. Res.}} \bibinfo{volume}{26},
  pp. \bibinfo{pages}{35--99}, \doi{10.1613/jair.1869}.

\bibitemdeclare{article}{Buchi1960}
\bibitem{Buchi1960}
\bibinfo{author}{J.~\surnamestart Büchi\surnameend} (\bibinfo{year}{1960}):
  \emph{\bibinfo{title}{{Weak Second-Order Arithmetic and Finite Automata}}}.
\newblock {\sl \bibinfo{journal}{Mathematical Logic Quarterly - MLQ}}
  \bibinfo{volume}{6}, pp. \bibinfo{pages}{66--92},
  \doi{10.1002/malq.19600060105}.

\bibitemdeclare{inproceedings}{CamachoBMM18}
\bibitem{CamachoBMM18}
\bibinfo{author}{Alberto \surnamestart Camacho\surnameend},
  \bibinfo{author}{Jorge~A. \surnamestart Baier\surnameend},
  \bibinfo{author}{Christian~J. \surnamestart Muise\surnameend} \&
  \bibinfo{author}{Sheila~A. \surnamestart McIlraith\surnameend}
  (\bibinfo{year}{2018}): \emph{\bibinfo{title}{{Finite {LTL} Synthesis as
  Planning}}}.
\newblock In: {\sl \bibinfo{booktitle}{{ICAPS}}}, pp. \bibinfo{pages}{29--38}.

\bibitemdeclare{inproceedings}{CamachoBM19}
\bibitem{CamachoBM19}
\bibinfo{author}{Alberto \surnamestart Camacho\surnameend},
  \bibinfo{author}{Meghyn \surnamestart Bienvenu\surnameend} \&
  \bibinfo{author}{Sheila~A. \surnamestart McIlraith\surnameend}
  (\bibinfo{year}{2019}): \emph{\bibinfo{title}{{Towards a Unified View of {AI}
  Planning and Reactive Synthesis}}}.
\newblock In: {\sl \bibinfo{booktitle}{{ICAPS}}}, pp. \bibinfo{pages}{58--67}.

\bibitemdeclare{article}{ChandraKS81}
\bibitem{ChandraKS81}
\bibinfo{author}{Ashok~K. \surnamestart Chandra\surnameend},
  \bibinfo{author}{Dexter \surnamestart Kozen\surnameend} \&
  \bibinfo{author}{Larry~J. \surnamestart Stockmeyer\surnameend}
  (\bibinfo{year}{1981}): \emph{\bibinfo{title}{Alternation}}.
\newblock {\sl \bibinfo{journal}{J. {ACM}}}
  \bibinfo{volume}{28}(\bibinfo{number}{1}), pp. \bibinfo{pages}{114--133},
  \doi{10.1145/322234.322243}.

\bibitemdeclare{inproceedings}{GiacomoV13}
\bibitem{GiacomoV13}
\bibinfo{author}{Giuseppe \surnamestart {De Giacomo}\surnameend} \&
  \bibinfo{author}{Moshe~Y. \surnamestart Vardi\surnameend}
  (\bibinfo{year}{2013}): \emph{\bibinfo{title}{{Linear Temporal Logic and
  Linear Dynamic Logic on Finite Traces}}}.
\newblock In: {\sl \bibinfo{booktitle}{{IJCAI}}}, pp.
  \bibinfo{pages}{854--860}.

\bibitemdeclare{inproceedings}{GiacomoV15}
\bibitem{GiacomoV15}
\bibinfo{author}{Giuseppe \surnamestart {De Giacomo}\surnameend} \&
  \bibinfo{author}{Moshe~Y. \surnamestart Vardi\surnameend}
  (\bibinfo{year}{2015}): \emph{\bibinfo{title}{{Synthesis for {LTL} and {LDL}
  on Finite Traces}}}.
\newblock In: {\sl \bibinfo{booktitle}{{IJCAI}}}, pp.
  \bibinfo{pages}{1558--1564}.

\bibitemdeclare{inproceedings}{GiacomoV16}
\bibitem{GiacomoV16}
\bibinfo{author}{Giuseppe \surnamestart {De Giacomo}\surnameend} \&
  \bibinfo{author}{Moshe~Y. \surnamestart Vardi\surnameend}
  (\bibinfo{year}{2016}): \emph{\bibinfo{title}{{LTL\({}_{\mbox{f}}\) and
  LDL\({}_{\mbox{f}}\) Synthesis under Partial Observability}}}.
\newblock In: {\sl \bibinfo{booktitle}{{IJCAI}}}, pp.
  \bibinfo{pages}{1044--1050}.

\bibitemdeclare{inbook}{DoyenR11}
\bibitem{DoyenR11}
\bibinfo{author}{Laurent \surnamestart Doyen\surnameend} \&
  \bibinfo{author}{Jean-François \surnamestart Raskin\surnameend}
  (\bibinfo{year}{2011}): \emph{\bibinfo{title}{Games with Imperfect
  Information: Theory and Algorithms}}, p. \bibinfo{pages}{185–212}.
\newblock \bibinfo{publisher}{Cambridge University Press},
  \doi{10.1017/CBO9780511973468.007}.

\bibitemdeclare{inproceedings}{spot}
\bibitem{spot}
\bibinfo{author}{Alexandre \surnamestart Duret-Lutz\surnameend},
  \bibinfo{author}{Alexandre \surnamestart Lewkowicz\surnameend},
  \bibinfo{author}{Amaury \surnamestart Fauchille\surnameend},
  \bibinfo{author}{Thibaud \surnamestart Michaud\surnameend},
  \bibinfo{author}{Etienne \surnamestart Renault\surnameend} \&
  \bibinfo{author}{Laurent \surnamestart Xu\surnameend} (\bibinfo{year}{2016}):
  \emph{\bibinfo{title}{{Spot 2.0 --- A Framework for {LTL} and
  $\omega$-automata Manipulation}}}.
\newblock In: {\sl \bibinfo{booktitle}{ATVA}},
  \doi{10.1007/978-3-319-46520-3\_8}.

\bibitemdeclare{inproceedings}{Ehlers11}
\bibitem{Ehlers11}
\bibinfo{author}{R{\"{u}}diger \surnamestart Ehlers\surnameend}
  (\bibinfo{year}{2011}): \emph{\bibinfo{title}{{Unbeast: Symbolic Bounded
  Synthesis}}}.
\newblock In \bibinfo{editor}{Parosh~Aziz \surnamestart Abdulla\surnameend} \&
  \bibinfo{editor}{K.~Rustan~M. \surnamestart Leino\surnameend}, editors: {\sl
  \bibinfo{booktitle}{{TACAS}}}, {\sl \bibinfo{series}{Lecture Notes in
  Computer Science}} \bibinfo{volume}{6605}, \bibinfo{publisher}{Springer}, pp.
  \bibinfo{pages}{272--275}, \doi{10.1007/978-3-642-19835-9\_25}.

\bibitemdeclare{inproceedings}{FaymonvilleFT17}
\bibitem{FaymonvilleFT17}
\bibinfo{author}{Peter \surnamestart Faymonville\surnameend},
  \bibinfo{author}{Bernd \surnamestart Finkbeiner\surnameend} \&
  \bibinfo{author}{Leander \surnamestart Tentrup\surnameend}
  (\bibinfo{year}{2017}): \emph{\bibinfo{title}{{BoSy: An Experimentation
  Framework for Bounded Synthesis}}}.
\newblock In: {\sl \bibinfo{booktitle}{{CAV}}}, pp. \bibinfo{pages}{325--332},
  \doi{10.1007/978-3-319-63390-9\_17}.

\bibitemdeclare{inproceedings}{FeltYBS93}
\bibitem{FeltYBS93}
\bibinfo{author}{Eric \surnamestart Felt\surnameend}, \bibinfo{author}{Gary
  \surnamestart York\surnameend}, \bibinfo{author}{Robert~K. \surnamestart
  Brayton\surnameend} \& \bibinfo{author}{Alberto~L. \surnamestart
  Sangiovanni{-}Vincentelli\surnameend} (\bibinfo{year}{1993}):
  \emph{\bibinfo{title}{{Dynamic Variable Reordering for {BDD} Minimization}}}.
\newblock In: {\sl \bibinfo{booktitle}{{EURO-DAC}}}, \bibinfo{publisher}{{IEEE}
  Computer Society}, pp. \bibinfo{pages}{130--135},
  \doi{10.1109/EURDAC.1993.410627}.

\bibitemdeclare{article}{FislerV02}
\bibitem{FislerV02}
\bibinfo{author}{Kathi \surnamestart Fisler\surnameend} \&
  \bibinfo{author}{Moshe~Y. \surnamestart Vardi\surnameend}
  (\bibinfo{year}{2002}): \emph{\bibinfo{title}{Bisimulation Minimization and
  Symbolic Model Checking}}.
\newblock {\sl \bibinfo{journal}{Formal Methods Syst. Des.}}
  \bibinfo{volume}{21}(\bibinfo{number}{1}), pp. \bibinfo{pages}{39--78},
  \doi{10.1023/A:1016091902809}.

\bibitemdeclare{inproceedings}{FogartyKVW13}
\bibitem{FogartyKVW13}
\bibinfo{author}{Seth \surnamestart Fogarty\surnameend}, \bibinfo{author}{Orna
  \surnamestart Kupferman\surnameend}, \bibinfo{author}{Moshe~Y. \surnamestart
  Vardi\surnameend} \& \bibinfo{author}{Thomas \surnamestart Wilke\surnameend}
  (\bibinfo{year}{2013}): \emph{\bibinfo{title}{{Profile Trees for B{\"{u}}chi
  Word Automata, with Application to Determinization}}}.
\newblock In: {\sl \bibinfo{booktitle}{{GandALF}}}, pp.
  \bibinfo{pages}{107--121}, \doi{10.4204/EPTCS.119.11}.

\bibitemdeclare{inproceedings}{GoldmanB96}
\bibitem{GoldmanB96}
\bibinfo{author}{Robert~P. \surnamestart Goldman\surnameend} \&
  \bibinfo{author}{Mark~S. \surnamestart Boddy\surnameend}
  (\bibinfo{year}{1996}): \emph{\bibinfo{title}{{Expressive Planning and
  Explicit Knowledge}}}.
\newblock In: {\sl \bibinfo{booktitle}{Proceedings of the Third International
  Conference on Artificial Intelligence Planning Systems}}, pp.
  \bibinfo{pages}{110--117}.

\bibitemdeclare{inproceedings}{HeWKV19}
\bibitem{HeWKV19}
\bibinfo{author}{Keliang \surnamestart He\surnameend},
  \bibinfo{author}{Andrew~M. \surnamestart Wells\surnameend},
  \bibinfo{author}{Lydia~E. \surnamestart Kavraki\surnameend} \&
  \bibinfo{author}{Moshe~Y. \surnamestart Vardi\surnameend}
  (\bibinfo{year}{2019}): \emph{\bibinfo{title}{{Efficient Symbolic Reactive
  Synthesis for Finite-Horizon Tasks}}}.
\newblock In: {\sl \bibinfo{booktitle}{{ICRA}}}, pp.
  \bibinfo{pages}{8993--8999}, \doi{10.1109/ICRA.2019.8794170}.

\bibitemdeclare{inproceedings}{HenriksenJJKPRS95}
\bibitem{HenriksenJJKPRS95}
\bibinfo{author}{Jesper~G. \surnamestart Henriksen\surnameend},
  \bibinfo{author}{Jakob~L. \surnamestart Jensen\surnameend},
  \bibinfo{author}{Michael~E. \surnamestart J{\o}rgensen\surnameend},
  \bibinfo{author}{Nils \surnamestart Klarlund\surnameend},
  \bibinfo{author}{Robert \surnamestart Paige\surnameend},
  \bibinfo{author}{Theis \surnamestart Rauhe\surnameend} \&
  \bibinfo{author}{Anders \surnamestart Sandholm\surnameend}
  (\bibinfo{year}{1995}): \emph{\bibinfo{title}{{Mona: Monadic Second-Order
  Logic in Practice}}}.
\newblock In: {\sl \bibinfo{booktitle}{{TACAS}}}, pp. \bibinfo{pages}{89--110},
  \doi{10.1007/3-540-60630-0\_5}.

\bibitemdeclare{inproceedings}{KupfermanV97}
\bibitem{KupfermanV97}
\bibinfo{author}{Orna \surnamestart Kupferman\surnameend} \&
  \bibinfo{author}{Moshe \surnamestart Vardi\surnameend}
  (\bibinfo{year}{1997}): \emph{\bibinfo{title}{{Synthesis with Incomplete
  Informatio}}}.
\newblock In: {\sl \bibinfo{booktitle}{{ICTL}}}, pp.
  \bibinfo{pages}{1044--1050}, \doi{10.1007/978-94-015-9586-5_6}.

\bibitemdeclare{inproceedings}{MaliahBKS14}
\bibitem{MaliahBKS14}
\bibinfo{author}{Shlomi \surnamestart Maliah\surnameend},
  \bibinfo{author}{Ronen~I. \surnamestart Brafman\surnameend},
  \bibinfo{author}{Erez \surnamestart Karpas\surnameend} \&
  \bibinfo{author}{Guy \surnamestart Shani\surnameend} (\bibinfo{year}{2014}):
  \emph{\bibinfo{title}{{Partially Observable Online Contingent Planning Using
  Landmark Heuristics}}}.
\newblock In: {\sl \bibinfo{booktitle}{{ICAPS}}}.

\bibitemdeclare{inproceedings}{MeyerSL18}
\bibitem{MeyerSL18}
\bibinfo{author}{Philipp~J. \surnamestart Meyer\surnameend},
  \bibinfo{author}{Salomon \surnamestart Sickert\surnameend} \&
  \bibinfo{author}{Michael \surnamestart Luttenberger\surnameend}
  (\bibinfo{year}{2018}): \emph{\bibinfo{title}{{Strix: Explicit Reactive
  Synthesis Strikes Back!}}}
\newblock In \bibinfo{editor}{Hana \surnamestart Chockler\surnameend} \&
  \bibinfo{editor}{Georg \surnamestart Weissenbacher\surnameend}, editors: {\sl
  \bibinfo{booktitle}{{CAV}}}, {\sl \bibinfo{series}{Lecture Notes in Computer
  Science}} \bibinfo{volume}{10981}, \bibinfo{publisher}{Springer}, pp.
  \bibinfo{pages}{578--586}, \doi{10.1007/978-3-319-96145-3\_31}.

\bibitemdeclare{inproceedings}{Pnueli77}
\bibitem{Pnueli77}
\bibinfo{author}{Amir \surnamestart Pnueli\surnameend} (\bibinfo{year}{1977}):
  \emph{\bibinfo{title}{{The Temporal Logic of Programs}}}.
\newblock In: {\sl \bibinfo{booktitle}{18th Annual Symposium on Foundations of
  Computer Science}}, pp. \bibinfo{pages}{46--57}, \doi{10.1109/SFCS.1977.32}.

\bibitemdeclare{inproceedings}{PnueliR89}
\bibitem{PnueliR89}
\bibinfo{author}{Amir \surnamestart Pnueli\surnameend} \& \bibinfo{author}{Roni
  \surnamestart Rosner\surnameend} (\bibinfo{year}{1989}):
  \emph{\bibinfo{title}{{On the Synthesis of a Reactive Module}}}.
\newblock In: {\sl \bibinfo{booktitle}{Sixteenth Annual {ACM} Symposium on
  Principles of Programming Languages}}, pp. \bibinfo{pages}{179--190},
  \doi{10.1145/75277.75293}.

\bibitemdeclare{article}{Reif84}
\bibitem{Reif84}
\bibinfo{author}{John~H. \surnamestart Reif\surnameend} (\bibinfo{year}{1984}):
  \emph{\bibinfo{title}{{The Complexity of Two-Player Games of Incomplete
  Information}}}.
\newblock {\sl \bibinfo{journal}{J. Comput. Syst. Sci.}}
  \bibinfo{volume}{29}(\bibinfo{number}{2}), pp. \bibinfo{pages}{274--301},
  \doi{10.1016/0022-0000(84)90034-5}.

\bibitemdeclare{inproceedings}{Rintanen04a}
\bibitem{Rintanen04a}
\bibinfo{author}{Jussi \surnamestart Rintanen\surnameend}
  (\bibinfo{year}{2004}): \emph{\bibinfo{title}{{Complexity of Planning with
  Partial Observability}}}.
\newblock In: {\sl \bibinfo{booktitle}{{ICAPS}}}, pp.
  \bibinfo{pages}{345--354}.

\bibitemdeclare{article}{Rosner91}
\bibitem{Rosner91}
\bibinfo{author}{Roni \surnamestart Rosner\surnameend} (\bibinfo{year}{1991}):
  \emph{\bibinfo{title}{{Modular Synthesis of Reactive Systems}}}.

\bibitemdeclare{inproceedings}{SistlaVW85}
\bibitem{SistlaVW85}
\bibinfo{author}{A.~Prasad \surnamestart Sistla\surnameend},
  \bibinfo{author}{Moshe~Y. \surnamestart Vardi\surnameend} \&
  \bibinfo{author}{Pierre \surnamestart Wolper\surnameend}
  (\bibinfo{year}{1985}): \emph{\bibinfo{title}{{The Complementation Problem
  for B{\"{u}}chi Automata with Applications to Temporal Logic (Extended
  Abstract)}}}.
\newblock In: {\sl \bibinfo{booktitle}{Automata, Languages and Programming,
  12th Colloquium}}, pp. \bibinfo{pages}{465--474}, \doi{10.1007/BFb0015772}.

\bibitemdeclare{article}{StockmeyerC79}
\bibitem{StockmeyerC79}
\bibinfo{author}{Larry~J. \surnamestart Stockmeyer\surnameend} \&
  \bibinfo{author}{Ashok~K. \surnamestart Chandra\surnameend}
  (\bibinfo{year}{1979}): \emph{\bibinfo{title}{{Provably Difficult
  Combinatorial Games}}}.
\newblock {\sl \bibinfo{journal}{{SIAM} J. Comput.}}
  \bibinfo{volume}{8}(\bibinfo{number}{2}), pp. \bibinfo{pages}{151--174},
  \doi{10.1137/0208013}.

\bibitemdeclare{article}{TabakovRV12}
\bibitem{TabakovRV12}
\bibinfo{author}{Deian \surnamestart Tabakov\surnameend},
  \bibinfo{author}{Kristin~Y. \surnamestart Rozier\surnameend} \&
  \bibinfo{author}{Moshe~Y. \surnamestart Vardi\surnameend}
  (\bibinfo{year}{2012}): \emph{\bibinfo{title}{{Optimized Temporal Monitors
  for SystemC}}}.
\newblock {\sl \bibinfo{journal}{Formal Methods in System Design}}
  \bibinfo{volume}{41}(\bibinfo{number}{3}), pp. \bibinfo{pages}{236--268},
  \doi{10.1007/s10703-011-0139-8}.

\bibitemdeclare{inproceedings}{VardiS85}
\bibitem{VardiS85}
\bibinfo{author}{Moshe~Y. \surnamestart Vardi\surnameend} \&
  \bibinfo{author}{Larry~J. \surnamestart Stockmeyer\surnameend}
  (\bibinfo{year}{1985}): \emph{\bibinfo{title}{{Improved Upper and Lower
  Bounds for Modal Logics of Programs: Preliminary Report}}}.
\newblock In: {\sl \bibinfo{booktitle}{Proceedings of the 17th Annual {ACM}
  Symposium on Theory of Computing}}, pp. \bibinfo{pages}{240--251},
  \doi{10.1145/22145.22173}.

\bibitemdeclare{inproceedings}{ZhuPV19}
\bibitem{ZhuPV19}
\bibinfo{author}{Shufang \surnamestart Zhu\surnameend},
  \bibinfo{author}{Geguang \surnamestart Pu\surnameend} \&
  \bibinfo{author}{Moshe~Y. \surnamestart Vardi\surnameend}
  (\bibinfo{year}{2019}): \emph{\bibinfo{title}{{First-Order vs. Second-Order
  Encodings for LTL$_f$-to-Automata Translation}}}.
\newblock In: {\sl \bibinfo{booktitle}{{TAMC}}}, pp. \bibinfo{pages}{684--705},
  \doi{10.1007/978-3-030-14812-6\_43}.

\bibitemdeclare{inproceedings}{ZhuTLPV17}
\bibitem{ZhuTLPV17}
\bibinfo{author}{Shufang \surnamestart Zhu\surnameend},
  \bibinfo{author}{Lucas~M. \surnamestart Tabajara\surnameend},
  \bibinfo{author}{Jianwen \surnamestart Li\surnameend},
  \bibinfo{author}{Geguang \surnamestart Pu\surnameend} \&
  \bibinfo{author}{Moshe~Y. \surnamestart Vardi\surnameend}
  (\bibinfo{year}{2017}): \emph{\bibinfo{title}{{Symbolic LTLf Synthesis}}}.
\newblock In: {\sl \bibinfo{booktitle}{{IJCAI}}}, pp.
  \bibinfo{pages}{1362--1369}, \doi{10.24963/ijcai.2017/189}.

\bibitemdeclare{article}{Zielonka98}
\bibitem{Zielonka98}
\bibinfo{author}{Wieslaw \surnamestart Zielonka\surnameend}
  (\bibinfo{year}{1998}): \emph{\bibinfo{title}{{Infinite Games on Finitely
  Coloured Graphs with Applications to Automata on Infinite Trees}}}.
\newblock {\sl \bibinfo{journal}{Theor. Comput. Sci.}}
  \bibinfo{volume}{200}(\bibinfo{number}{1-2}), pp. \bibinfo{pages}{135--183},
  \doi{10.1016/S0304-3975(98)00009-7}.

\end{thebibliography}
\end{document}